\documentclass[a4paper, amsfonts, amssymb, amsmath, preprint, showkeys, nofootinbib, twoside]{revtex4-1}
\usepackage[english]{babel}
\usepackage[utf8]{inputenc}
\usepackage{feynmp-auto}
\usepackage{mathtools}
\usepackage{gensymb}
\usepackage{slashed}
\usepackage[colorinlistoftodos, color=green!40, prependcaption]{todonotes}
\usepackage{amsthm}
\usepackage{mathtools}
\usepackage{physics}
\usepackage{xcolor}
\usepackage{graphicx}
\usepackage[left=23mm,right=13mm,top=35mm,columnsep=15pt]{geometry} 
\usepackage{adjustbox}
\usepackage{placeins}
\usepackage[T1]{fontenc}
\usepackage{lipsum}
\usepackage{csquotes}
\usepackage[pdftex, pdftitle={Article}, pdfauthor={Author}]{hyperref} 

\begin{document}
\title{On the Gravitational Wave Counterpart to a Gamma-ray Galactic Center Signal from Millisecond Pulsars}

\author{Kayla Bartel}
\email[Correspondence email address: ]{kmbartel@ucsc.edu}
\affiliation{Department of Physics, University of California, Santa Cruz (UCSC),
Santa Cruz, CA 95064, USA}

\author{Stefano Profumo}
    \email{profumo@ucsc.edu}
\affiliation{Department of Physics, University of California, Santa Cruz (UCSC),
Santa Cruz, CA 95064, USA}
\affiliation{Santa Cruz Institute for Particle Physics (SCIPP),
Santa Cruz, CA 95064, USA}

\date{\today} 

\begin{abstract}
The new tools of gravitational wave and multi-messenger
astronomy allow for the study of astrophysical phenomenon in new ways and enables light to be shed on some of the longest-enduring mysteries of high-energy astrophysics. Among the latter stands the Galactic center gamma-ray excess, associated with a source whose nature could be annihilating dark matter or a yet-unresolved population of millisecond pulsars (MSPs). MSPs are most likely asymmetric about their axis of rotation, and are thus thought to also source quasi-monochromatic gravitational waves, that dark matter processes would not emit. Using statistical methods, we simulate realistic MSP population samples with differing morphology and moment of inertia, that could give rise to the gamma-ray excess, and we compute the corresponding gravitational wave signal amplitude and frequency. We find that the gravitational wave signal frequency likely ranges between $\sim$200 and 1400 Hz, and that the collective dimensionless strain from the center of the Galaxy has an amplitude between $10^{-26}$ and $10^{-24}$, thus most likely beyond current and near-term detectors, unless the unresolved MSPs are extraordinarily gamma-ray dim. 
\end{abstract}


\maketitle

\section{Introduction} \label{sec:introduction}

In recent years, we have entered an age of multi-messenger astrophysics. This is especially apparent in the field of gravitational wave astronomy, which now allows researchers to probe new phenomena and old mysteries alike with entirely new information. One of the most enduring astrophysical puzzles is the nature of the so-called Galactic Center Excess (GCE) \cite{goodenough2009possibleevidencedarkmatter}, a high-energy (1-100 GeV) gamma-ray signal from the center of the Milky Way Galaxy, discovered with data from the Large Area Telescope (LAT) on board the Fermi Gamma-ray Space Telescope \cite{goodenough2009possibleevidencedarkmatter, Slatyer_2022}. Different background models suggest that the GCE is either spherical or bulge-like in shape, with each morphology corresponding to a different potential origin \cite{Slatyer_2022}.

When first reported, the gamma-ray signal of the GCE was found to be consistent with that expected from annihilating dark matter (DM) with a DM density profile $\rho(r)$ resembling that of a Navarro-Frenk-White (NFW) profile \cite{Navarro:1996gj} with a radial dependence, in the inner regions, $\rho\sim r^{- \gamma}$, with $\gamma = 1.1$ \cite{goodenough2009possibleevidencedarkmatter}. Further studies \cite{Hooper_2011, Abazajian_2012, Gordon_2013, Daylan_2016} confirmed the resemblance to the square of a steepened NFW profile, and hence the gamma ray spectrum's spherical morphology, suggestive of a $\lesssim 100 $ GeV thermal relic DM annihilating to quark pairs \cite{Slatyer_2022}. However, it is impossible to rule out the possibility of other astrophysical point sources with a similar spectral shape and collective morphology, especially if the GCE has, instead, a bulge-like shape \cite{Slatyer_2022}. In fact, several authors have posited that the GCE is caused by an unresolved population of MSPs, with many models fitting the morphology and energetics of the GCE remarkably well \cite{Abazajian_2012, Eckner_2018, ploeg2021galactic}. 

MSPs are rapidly rotating neutron stars, with rotation periods in the milliseconds, whose rotational motion has been accelerated by accretion from a companion star \cite{MSPreview}. MSPs are found in the interstellar medium, as well as in globular clusters and galactic bulges \cite{Eckner_2018}. MSPs emit curvature radiation in the form of radio and gamma rays, driven by the motion of charged particles along strong magnetic field ($\sim 10^8 - 10^9$ G) lines. These emissions are not isotropic, with the strongest emission  along the magnetic axis -- reminiscent of a lighthouse \cite{ploeg2021galactic}. 

The predominant formation mechanism for MSPs in the Galactic center is believed to be associate with the so-called ``cycling'' formation scenario, in which a slowly rotating neutron star formed via a core-collapse supernova accretes material from a companion star during a low mass x-ray binary (LMXB) phase and is thus spun up into a MSP \cite{MSPreview, Gautam:2021wqn}. However, this formation mechanism has been found to predict too many LMXB systems in the Galactic center, casting doubt on its validity \cite{Gautam:2021wqn}. Recent work has instead suggested that a population of MSPs resulting from the accretion-induced collapse (AIC) of O-Ne white dwarfs in the Galactic bulge is a significantly more likely pathway -- both producing the correct intensity and morphology of the GCE signal while also remaining within LMXB constraints \cite{Gautam:2021wqn}. Additional work also suggests that individual MSPs, found to be approximately 20-27\% of MSPs in the Galactic center, are unlikely to contribute to the GCE signal; instead, binary systems are the likely culprit \cite{Macias_2019, Jiang:2019hal, Chanlaridis:2024rov}. The stellar encounter rate is significantly higher in the Galactic center than in the field, suggesting that the 20-27\% range could potentially be an empirical upper limit \cite{McTier2020}.

As noted above, there continues to be some ambiguity in the determination of the shape of the GCE, as it depends heavily on the model of the continuum gamma-ray backgrounds used. Both GALPROP \cite{Macias_2018, Macias_2019} and SkyFACT \cite{Storm_2017} suggest a bulge-like morphology, though due to uncertainty in current interstellar gas maps and the background models themselves, the ambiguity remains. Another method of determining the cause of the GCE is studying photon statistics: a DM signal would produce a smooth spatial distribution of flux while pulsars would produce more concentrated distributions \cite{Slatyer_2022}. Recent studies are conflicted, suggesting both that there is some evidence for point sources \cite{Buschmann_2020} and that the GCE is an asymmetric smooth emission \cite{Leane_2020, Leane_2020_2}.

Unlike DM annihilation, MSPs are slated to produce a gravitational wave signal; the latter originates from the non-spherical nature of the rotating neutron star, a consequence, in turn, of its large rotational speeds (see e.g. \cite{Agarwal_2022} and references therein). Thus, detecting the gravitational wave (GW) counterpart generated by MSPs could potentially solve the open question of the origin of the GCE, as pointed out in a few previous analyses \cite{Calore_2019, Agarwal_2022, Miller_2023, Carleo:2023qxu}.  For instance, in the present study, we re-assess the connection between the GCE and GW emission in the MSPs scenario for the GCE, utilizing state-of-the-art population models and statistical methods.

As we discuss below, we improve over, or significantly differ in our analysis from previous studies in a number of ways. For instance, Ref.~\cite{Calore_2019} assumes a single value for the MSPs ellipticities, and thus moment of inertia, location (all MSPs are assumed to be exactly at the GC), a simple statistical model for the frequency distribution, and a single population in the bulge; Ref.~\cite{Miller_2023} details the effects of different distributions in frequency and models for the moment of inertia/magnetic field distributions as well, but not in the spatial distribution; Ref.~\cite{Agarwal_2022} uses a single value for the moment of inertia, and a simplified model for the spatial distribution of MSPs, and focuses on the resulting anisotropic stochastic gravitational-wave background.

The remainder of this study is structured as follows: in Section \ref{MSP Models} and  \ref{Sampling the Millisecond Pulsar Population} we give an overview of the models and methods used to generate the MSPs population models and the resulting GW signal. Sections \ref{Ellipticity Distribution of Millisecond Pulsars}, \ref{Moment of Inertia Distribution for Millisecond Pulsars}, and \ref{Generating the Gravitational Wave Signal} contain discussion of the ellipticity and moment of inertia distribution of the population of MSPs under consideration and the impact on the generated GW signal, respectively. We end with remarks on the possibility of detection, our conclusions, and and outlook of future work in Sections \ref{Analysis and Possibility of Detection} and \ref{Conclusion}.

\section{Millisecond Pulsar Models} \label{MSP Models}
The theory of general relativity posits that gravitational waves -- perturbations in the fabric of space-time traveling at the speed of light -- exist and are detectable. Their existence has been conclusively supported by both direct and indirect observational evidence. These space-time distortions were in fact first directly observed via the detection of the binary star merger GW170817 by LIGO in 2017 \cite{Abbott_2017}. 

Due to conservation of mass and conservation of momentum, emission from the monopole and dipole is impossible \cite{Carroll:2004st}. Therefore, the lowest radiating multipole is the quadrupole \cite{Sieniawska_2021}.  In the case of MSPs, the systems of interest here, quasi-monochromatic gravitational waves can thus be emitted from asymmetric deformations of the pulsar mass distribution around the axis of rotation \cite{Agarwal_2022}. One likely origin of these deformations is a strong internal magnetic field, $B_{int}$, provided the field does not align with the axis of rotation \cite{Miller_2023}. The breaking of the axial asymmetry can be expressed in terms of the dimensionless equatorial ellipticity, $\epsilon$, which is proportional to the GW strain amplitude \cite{Agarwal_2022}. The ellipticity is defined in relation to the star's principle moments of inertia $I_{xx}, I_{yy}, I_{zz}$ 
 as \cite{Miller_2023}:

\begin{equation} \label{ellipticity_eqn}
    \epsilon = \frac{\left| I_{xx} - I_{yy} \right|}{I_{zz}}
\end{equation}
In most pulsar models  $|\epsilon| \ll 1$, $I_{xx} \simeq I_{yy} \simeq I_{zz} = I$, with a conservative estimate for $I$ being $1.1 \times 10^{38}$ kg ${\rm m}^2$ (see sec.~\ref{Moment of Inertia Distribution for Millisecond Pulsars} below for an extended discussion). Assuming a superconducting core, one can model the ellipticity as \cite{Miller_2023}:

\begin{equation} \label{Bfield_ellipticity_eqn}
    \epsilon \approx 10^{-8}\left(\frac{B_{int}}{10^{12} {\rm \ G}}\right).
\end{equation}
However, the internal magnetic field is not directly observable; Instead one must infer it from the MSP's external magnetic field, $B_{ext}$, which is estimated to be $B_{int} \simeq 150B_{ext}$ \cite{Miller_2023}. Note that this is a conservative estimate, as the internal magnetic field could be as much as $10^4$ times larger than the external magnetic field \cite{Miller_2023}.

Following Ref.~\cite{ploeg2021galactic}, here we model the probability density of the external magnetic field of a MSP with a log-normal distribution:

\begin{equation} \label{Bfield_prob_eqn}
    p(\log_{10}(B)|B_{med}, \sigma_B) = \frac{1}{\sqrt{2\pi}\sigma_B}\exp\left({-\frac{(\log_{10}(B) - \log_{10}(B_{med})^2}{2\sigma_B^2}}\right),
\end{equation}
with $\log_{10}(B_{med})$ = 8.21, with $B_{med}$ in G, and $\sigma_B$ = 0.21.

The gravitational wave strain amplitude $h_0$ is then given by \cite{Miller_2023}:

\begin{equation} \label{GW_amplitude_eqn}
    h_0 = \frac{16\pi^2G}{c^4}\frac{I_{zz}\epsilon f_{rot}^2}{d}
\end{equation}
with $d$ the distance to the MSP from Earth, $f_{rot}$ the MSP's rotational frequency, and $G$ the gravitational constant. These models assume that the ellipticities, and corresponding gravitational wave production, arise from isolated MSPs or MSPs with no orbital eccentricities--a simplifying assumption that we continue to adopt in the following work.

\section{Sampling the Millisecond Pulsar Population} \label{Sampling the Millisecond Pulsar Population}

To generate the GW signal corresponding to a population of MSPs associated with the GCE it is necessary to first generate a population of MSPs via Monte Carlo Markov Chain (MCMC) and inverse transform sampling from which a GW signal strength and morphology can be determined, as described below. Our benchmark for the population size is of 40,000, chosen to be within bounds given by Holst et. al. \cite{holst2024new} to appropriately recreate the GCE gamma-ray luminosity with a population of MSPs. We  also consider below, for comparison, a larger population, relevant if the intrinsic gamma-ray luminosity is suppressed compared to expectations.

\subsection{Millisecond Pulsar Densities in the Milky Way} \label{Millisecond Pulsar Densities in the Milky Way}

Motivated by semi-analytical fits to population synthesis models \cite{refpopsynth} we consider two possible MSP morphology, a ``Boxy'' morphology (consisting of several sub-components, as described below) and a ``Spherical'' one. This follows the analysis of Ref. \cite{Eckner_2018} and \cite{ploeg2021galactic} in which the observed GCE morphology and gamma ray luminosity was reproduced, lending credence to our choice of population models. As stated in Section \ref{sec:introduction}, the morphology of the GCE can be interpreted as either spherical or boxy depending on the specific background model used. Therefore we choose to model the MSP population taking only these two morphologies into account, leaving other possibilities for future work.

\begin{figure}[t]
    \centering
    \includegraphics[width=0.4\textwidth]{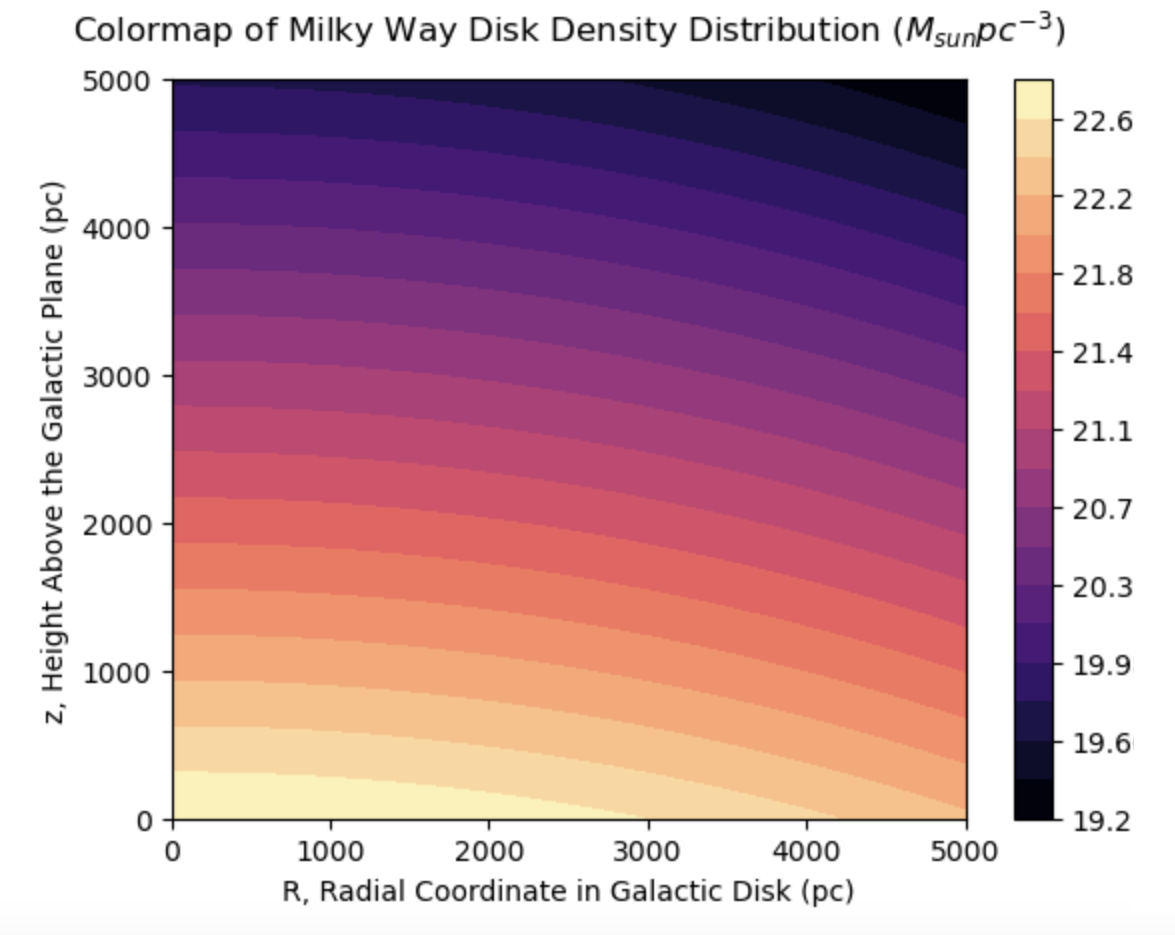}\qquad\includegraphics[width=0.44\textwidth]{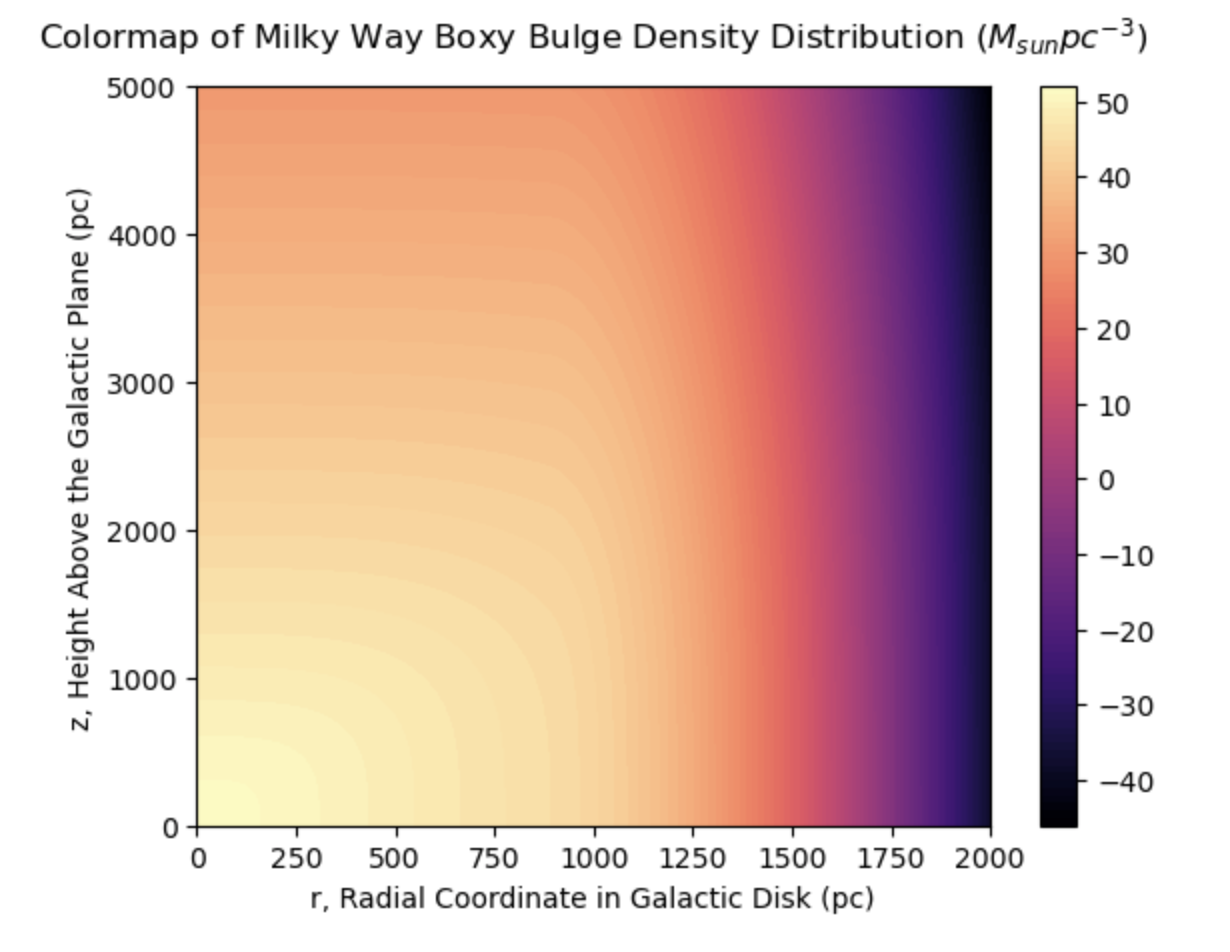}
    \caption{A colormap of the "Boxy" MSP density distribution in the Milky Way disk, in log10 scale; Left: the disk component; Right: the boxy bulge component.}
    \label{fig:rho_disk_colormap}\label{fig:rho_bb_colormap}
\end{figure}

The Boxy model includes two sub-populations: a disk  and a boxy bulge population, such that $\rho_{\text{boxy pop.}} = \rho_{\text{disk}} + \rho_{\text{boxy bulge}}$. The disk population, assumed to only depend on the Galactic latitude $z$ and radius $R$ coordinates, is assumed to follow a density of the form:

\begin{equation} \label{rho_disk_eqn}
    \rho_{\text{disk}}(R, z) = \frac{N_{\text{disk}}M\textsubscript{\(\odot\)}}{4\pi\sigma_r^2z_0}\exp{\left(\frac{-R^2}{2\sigma_r^2}\right)}\exp{\left(\frac{-|z|}{z_0}\right)},
\end{equation}

with $R^2 = x^2 + y^2$ the radial coordinate on the Galactic disk and $z$ the height above the Galactic plane \cite{ploeg2021galactic}. $\sigma_r$ and $z_0$ were chosen to be 4500 pc and 710 pc respectively in accordance with the findings of Ref.~\cite{ploeg2021galactic}. $N_{disk}$ is the number of MSPs in the Milky Way disk, and was chosen to be 6000 stars \cite{Yuan_2014}. See Fig.~\ref{fig:rho_disk_colormap}, left panel, for a plot of the MSP disk distribution.

The boxy bulge distribution is, following \cite{Ploeg_2020}, given by:

\begin{equation} \label{rho_bb_eqn}
    \rho_{\text{boxy bulge}}(R_{s}) \propto \text{sech}^2(R_{s}) \times \begin{cases} 
      1 & R\leq R_{\text{end}} \\
      \exp{-\frac{(R - R_{\text{end}})^2}{h_{\text{end}}^2}} & R > R_{\text{end}} \\ 
   \end{cases}
\end{equation}
with $R$ defined as above, $R_{\text{end}} = 3128$ pc, and $h_{\text{end}} = 461$ pc, and:  

\begin{equation} \label{R_perp_def_eqn}
    R_{\perp}^{C_{\perp}} = \left( \frac{|x'|}{1696\  \text{pc}} \right)^{C_{\perp}} + \left(\frac{|y'|}{642.6\  \text{pc}} \right)^{C_{\perp}}
\end{equation}

\begin{equation} \label{R_para_def_eqn}
    R_{S}^{C_{\parallel}} = R_{\perp}^{C_{\parallel}} + \left(\frac{|z'|}{442.5\  \text{pc}} \right)^{C_{\parallel}}
\end{equation}
where $C_{\parallel} = 3.501$ and $C_{\perp} = 1.574$. The coordinates $x^\prime,\ y^\prime$ and $z^\prime$ are Cartesian coordinates in the boxy bulge frame, which is rotated $13.79 \degree$ around the $z$-axis and then $0.023 \degree$ around the new $y$-axis, relative to the frame in which $x_{\odot} = -8300$ pc and $y_{\odot} = z_{\odot} = 0$. See Fig.~\ref{fig:rho_bb_colormap}, right, for a plot of the boxy bulge distribution.

\begin{figure}[t]
    \centering
    \includegraphics[width=9cm]{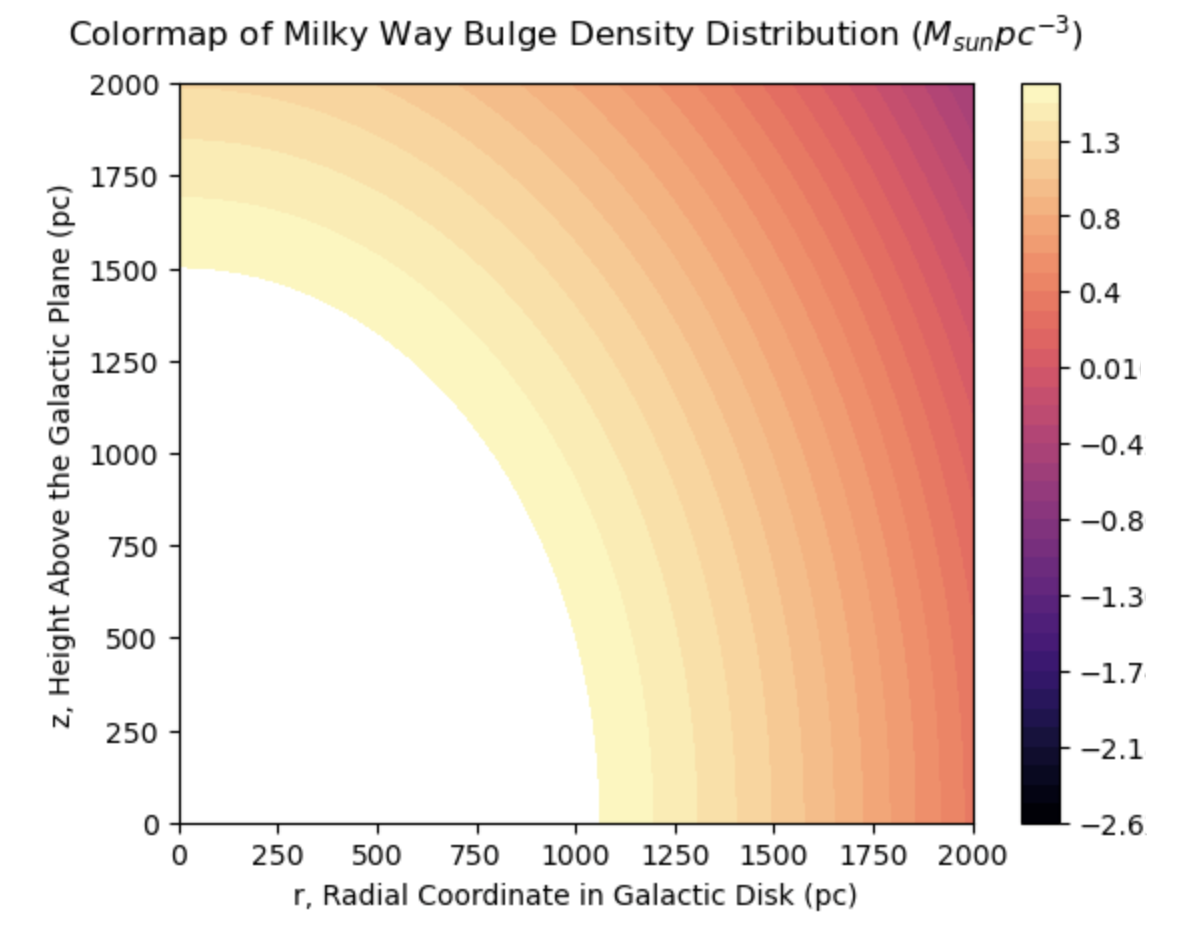}
    \caption{A colormap of the spherical MSP density distribution in the Milky Way in log10 scale.}
    \label{fig:rho_spherical_colormap}
\end{figure}

The second, spherical population of MSPs consists of $\rho_{\text{disk}}$ from Eq. \eqref{rho_disk_eqn} and a single, spherically symmetric density distribution, such that $\rho_{\text{spherical pop.}} = \rho_{\text{disk}} + \rho_{\text{sph}}$, given by:

\begin{equation} \label{rho_spherical_eqn}
    \rho_{\text{sph}}(R) = \rho_{0, sph} \frac{\exp{-R^2/R_m^2}}{\left( 1 + r/r_0\right)^{1.8}}
\end{equation}
with $R$ defined as above, $R_m = 1.9$ kpc, $R_0 = 100$ pc, and $\rho_{0, sph} = 45.27$ $M\textsubscript{\(\odot\)}\text{pc}^{-3}$ \cite{Eckner_2018}. See Fig.~\ref{fig:rho_spherical_colormap} for a plot of the spherical density distribution. 

To model the relevant parameters in the following work, two methods of generating samples were used: Monte Carlo Markov Chain sampling and Inverse Transform sampling. A typical Monte Carlo Markov Chain (MCMC) sampler undergoes an iterative procedure where, starting with an initial position, a new proposed position sampled from a transition distribution is accepted with some defined probability \cite{Foreman_Mackey_2013}. The sampler proposed by Goodman and Weare, and used here, is a significantly more efficient version, simultaneously evolving $N$ walkers in an ensemble to explore the parameter space and generate samples \cite{Foreman_Mackey_2013}. The proposal distribution for a single walker is based on the current positions of the remaining $N - 1$ walkers. To advance one walker to a new position, a walker is randomly chosen from the remaining pool and a new position is proposed and accepted according to the probability described in Ref.~\cite{Foreman_Mackey_2013}. This process is then repeated for each walker in the ensemble in series until a sample is generated \cite{Foreman_Mackey_2013}.

Sampling a 2D function, $f(x, y)$, requires splitting the function into two conditional functions, $f(x|y)$ and $f(y|x)$, as the MCMC method discussed above are 1-dimensional. For piece-wise functions f(x,y), the integral bounds for $f_x(x)$ and $f_y(y)$ are given by the piece-wise bounds.

\subsection{Sampling the Millisecond Pulsar Population} \label{Sampling the Pulsar Population Using MCMC Methods}

\begin{figure}[!t]
    \centering
    \includegraphics[width=7cm]{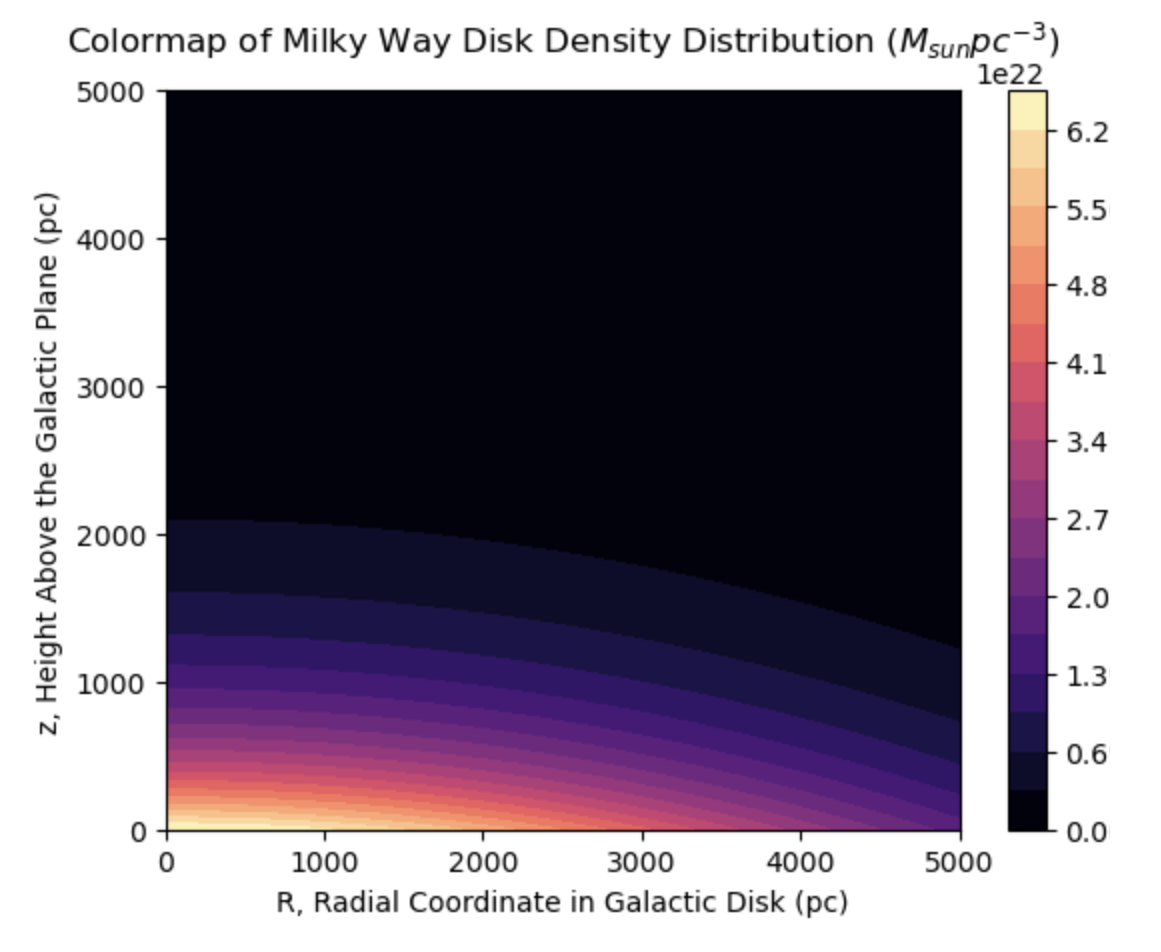}     \qquad
  \includegraphics[width=7cm]{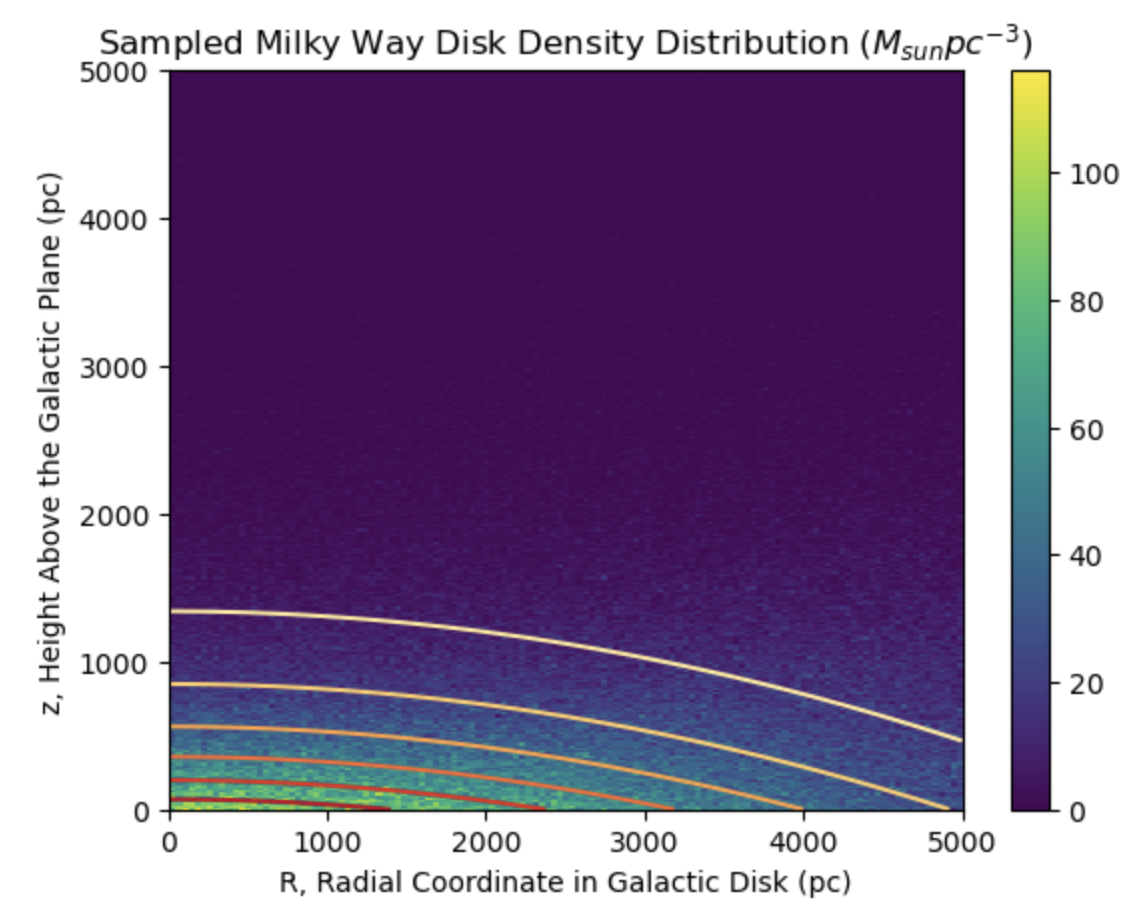}
    \caption{Comparison of the Milky Way Disk density and the corresponding sample generated by MCMC methods. The generated sampled of MSPs follows the same decay pattern and thus has a similar shape to the density distribution.}%
    \label{fig:MWD_sample_comparison}%
\end{figure}

To sample the Milky Way disk population and the spherical population, emcee, a MCMC package written in Python was used \cite{Foreman_Mackey_2013}. For the MW disk population, MCMC sampling was used to sample the corresponding conditional functions, given as:

\begin{equation} \label{MWD_Conditional_z_eqn}
    \rho(z|R) = \frac{1}{2 z_0}\exp{\left(\frac{-|z|}{z_0}\right)}
\end{equation}
and:

\begin{equation} \label{MWD_Conditional_R_eqn}
    \rho(R|z) = \frac{1}{2 \pi \sigma_R}\exp{\left(\frac{-R^2}{2 \sigma_R^2}\right)}
\end{equation}
with z, r, $z_0, \sigma_R$ defined as in Eq.~(\ref{rho_disk_eqn}). A comparison between the generated sample, plotted as a 2D histogram, and Eq.~(\ref{rho_disk_eqn}), plotted as a contour map, can be seen in Fig.~\ref{fig:MWD_sample_comparison}. To produce both boxy and the spherical bulge populations, Eq. ~(\ref{rho_bb_eqn}) and Eq.~(\ref{rho_spherical_eqn}) were also sampled via MCMC methods, and was confirmed to match with the density distribution pictured in Fig. ~\ref{fig:rho_bb_colormap} and Fig.~\ref{fig:rho_spherical_colormap}. 

\section{Ellipticity and Moment of Inertia Distribution} \label{Ellipticity Distribution of Millisecond Pulsars}\label{Moment of Inertia Distribution for Millisecond Pulsars}

\begin{figure}[t]
    \centering
    \includegraphics[width=8cm]{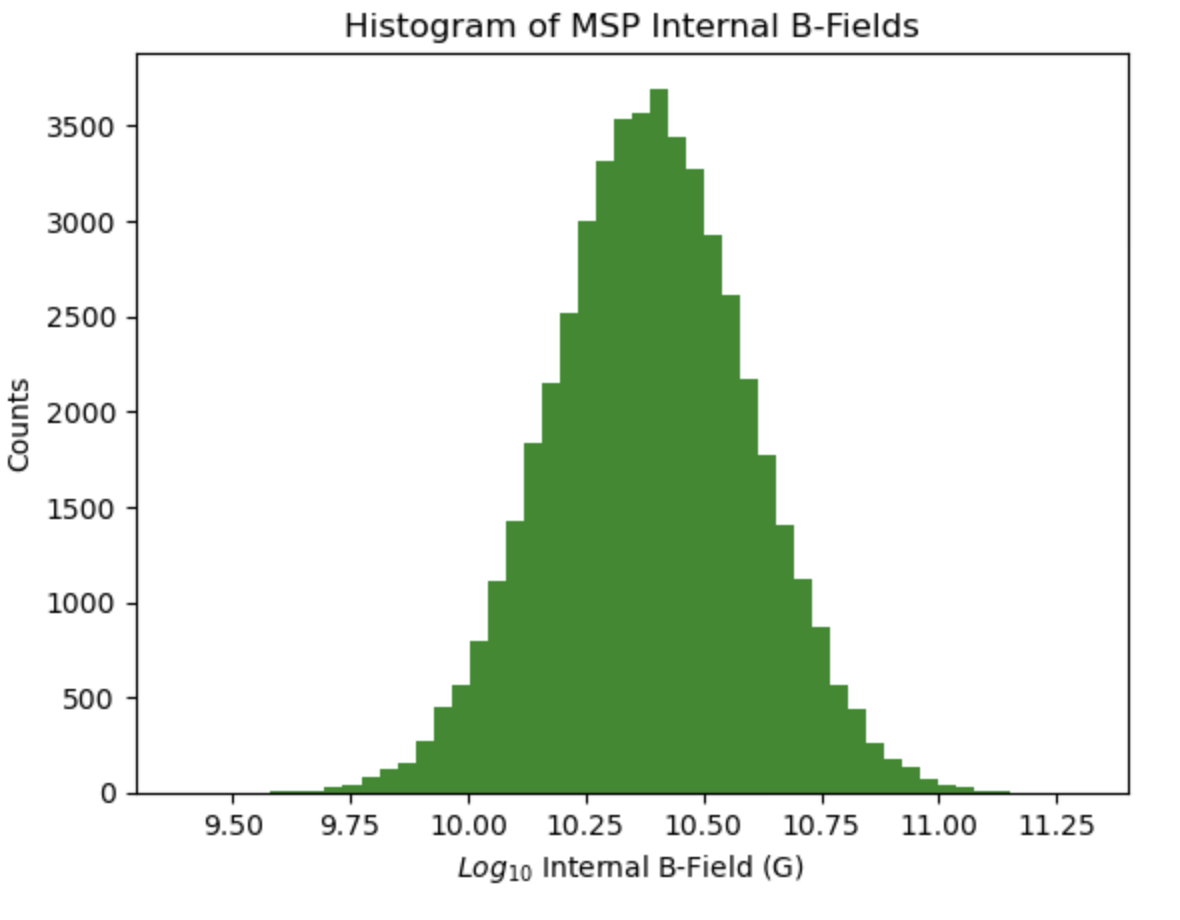}\qquad \includegraphics[width=8cm]{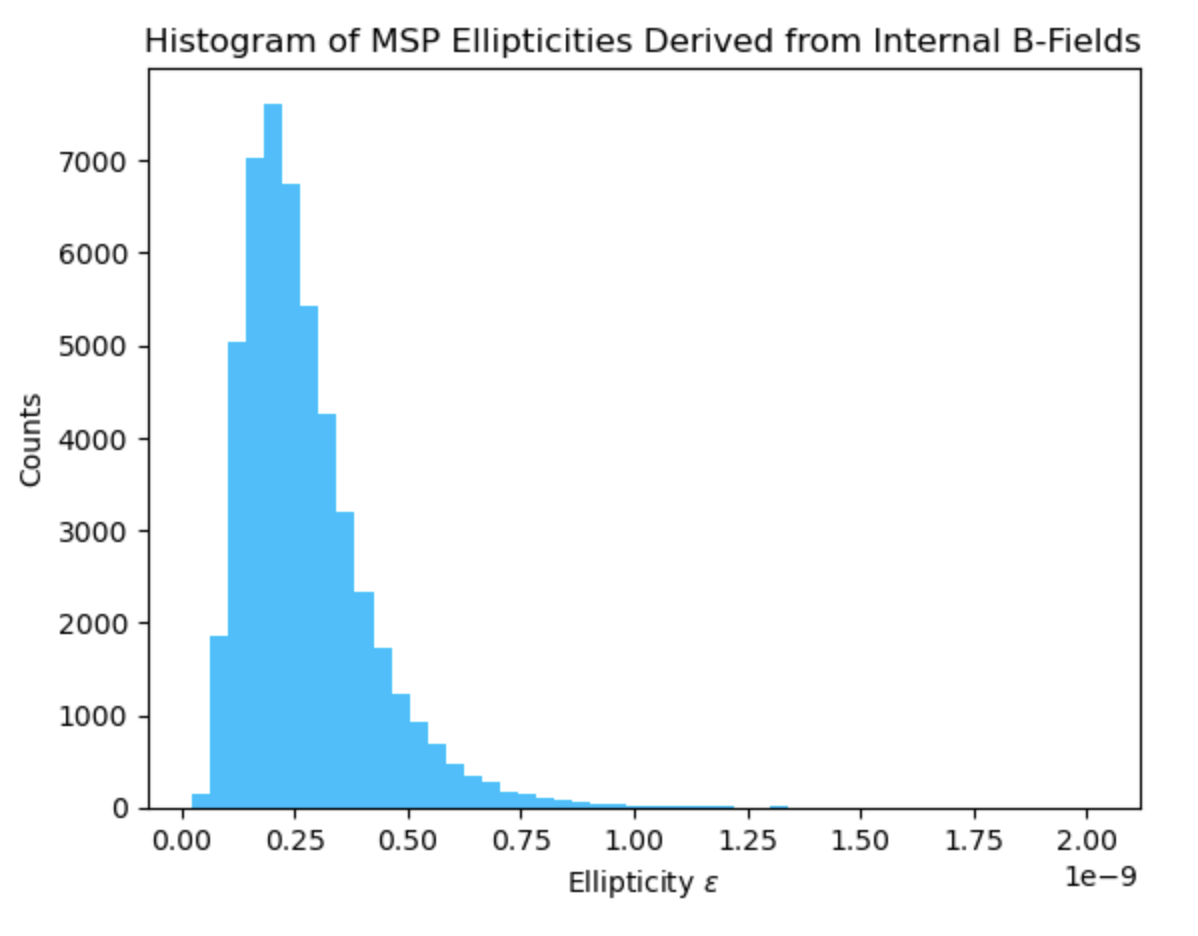}
    \caption{Left: A histogram of the internal magnetic fields of MSPs. Right: A histogram of the ellipticities of MSPs, derived from the internal magnetic field.}
    \label{fig:hist_int_b_field}\label{fig:ellipticity_int_b_field}
\end{figure}

To derive the ellipticity distribution of MSPs in the Milky Way, a method similar to Miller et al. \cite{Miller_2023} was used, with the probability distribution of the external magnetic fields given by Eq.~(\ref{Bfield_prob_eqn}) above. The corresponding internal magnetic fields were extracted from the external magnetic field distribution, assuming $B_{int} = 150B_{ext}$; we show the resulting magnetic field statistical distribution in Fig.~\ref{fig:hist_int_b_field}, left.

\begin{figure}[t]
    \centering
    \includegraphics[width=0.45\textwidth]{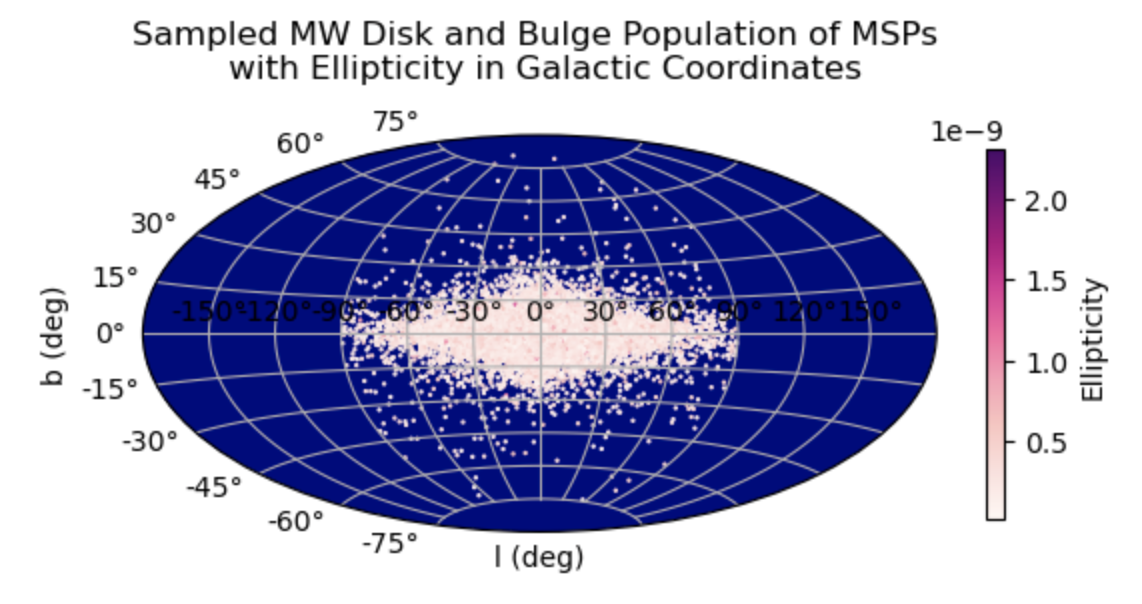}\qquad\includegraphics[width=0.45\textwidth]{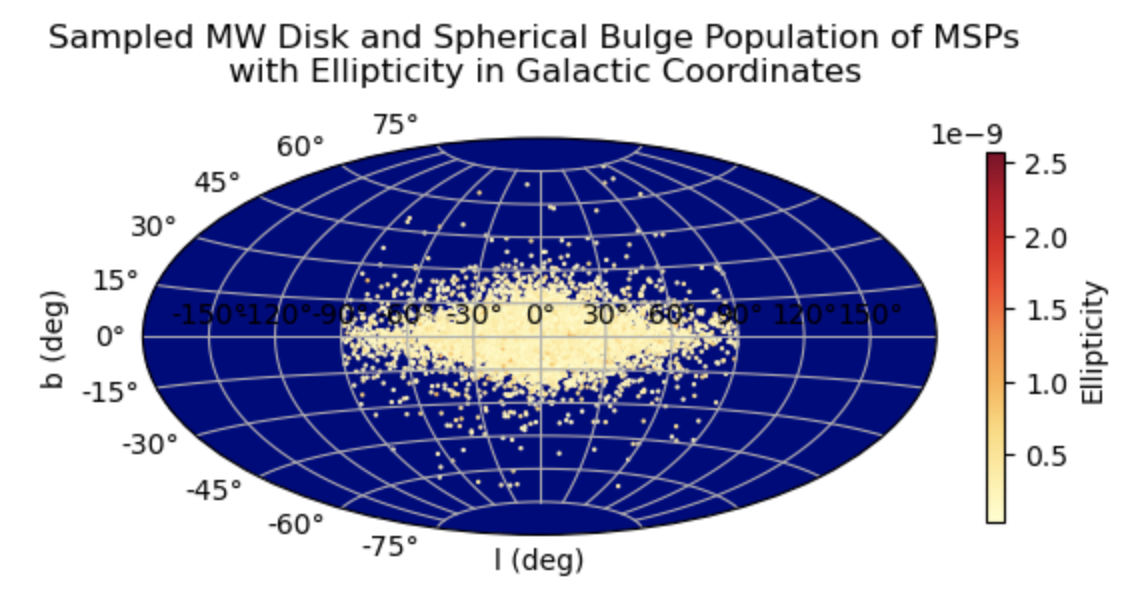}
    \caption{Locations and ellipticities of sampled MSPs for the boxy (left) and spherical (right) population models. The disk MSPs extend from -90 deg. to 90 deg. in the boxy population; the disk component is missing from the population model on the right, aimed at reproducing the possible GCE morphology.}
    \label{fig:MSP_aitoff_ellipticity}
\end{figure}

The ellipticity distribution was then folded with the morphological models described above, resulting in the projected pulsar ellipticity distribution shown in Fig.~\ref{fig:MSP_aitoff_ellipticity} for a randomly drawn sample of 40,000 MSPs for the Boxy (left) and Spherical (right) population models. The boxy distribution includes the Galactic disk component while the spherical distribution does not, resulting in the significant difference in morphology that can be observed. This is to better reflect the ambiguity of the GCE morphology as derived from the specific model of the gamma-ray backgrounds. As mentioned in Sec. \ref{sec:introduction}, due to uncertainties in the current interstellar gas maps and the background models used, a full determination of the morphology of the GCE has yet to be completely understood. Also note that the resultant ellipticity distribution is within the minimum ellipticity ranges described by Woan et al. \cite{Woan_2018} and Chen et al. \cite{Chen_2020} of $\epsilon \approx 10^{-9}$, and agrees with the values obtained by Miller et al. \cite{Miller_2023}, as well as within the recent constraints given by Holst et al. \cite{holst2024new}.

\begin{figure}[!t]
    \centering
    \includegraphics[width=0.5\textwidth]{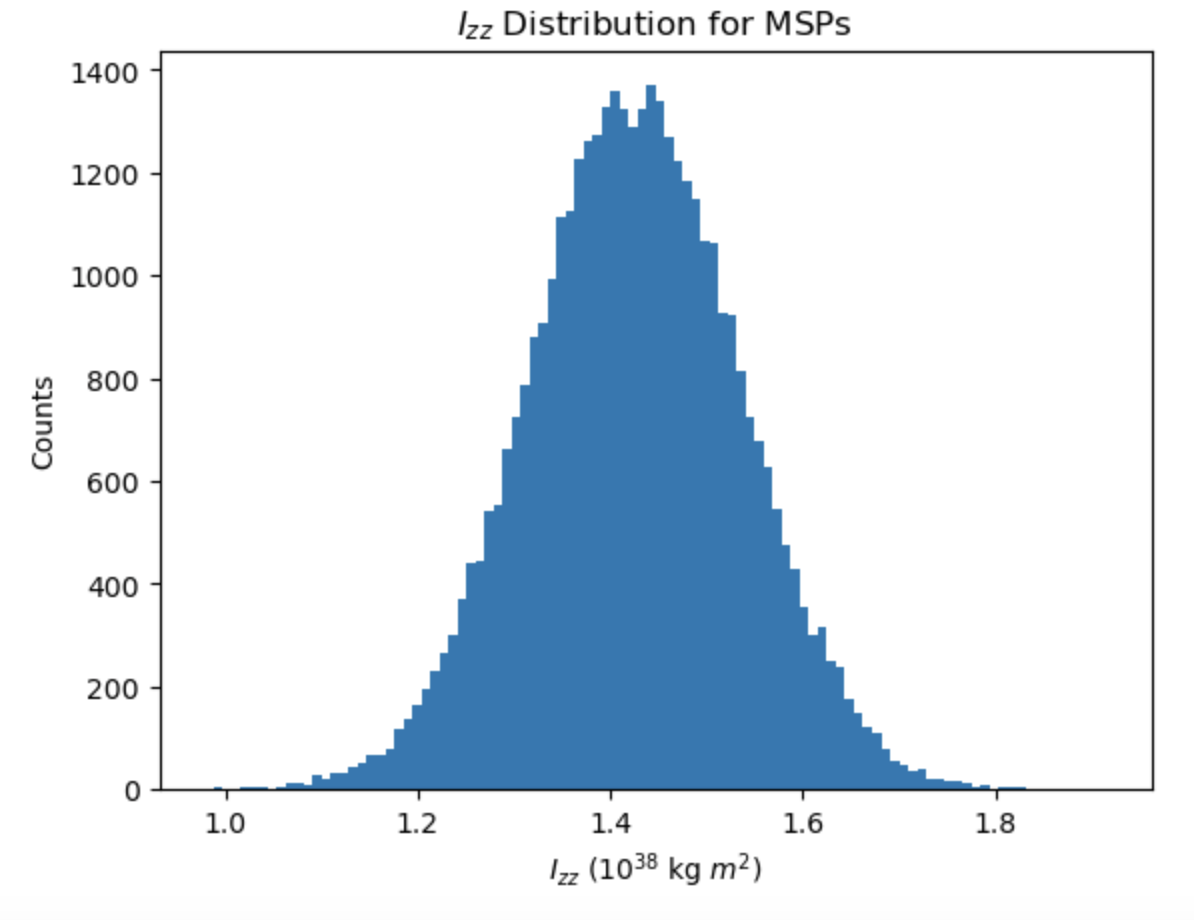}
    \caption{The distribution of moments of inertia for a MSP, with $\mu = 1.425$ and $\sigma = 0.110$.}
    \label{fig:MSP_I_zz_dist}
\end{figure}

The moment of inertia, $I_{zz}$, of our  MSP sample is determined by the currently largely unconstrained neutron star equation of state (EOS) \cite{worley2008nuclearconstraintsmomentainertia}. Different EOSs result in different $I_{zz}$ distributions for both different masses of neutron stars and different rotational frequencies \cite{worley2008nuclearconstraintsmomentainertia}. We assume that all MSPs are of ``canonical'' mass \cite{MSPreview}, of approximately 1.4 $M\textsubscript{\(\odot\)}$. Because the moment of inertia can range drastically depending upon the chosen EOS even for the same frequency value, as seen in Fig. 8 in Worley et al. \cite{worley2008nuclearconstraintsmomentainertia}, it was assumed that a normal distribution could be used for each MSP independent of rotational frequency. Therefore, unless sampled from a range between $10^{38}$ and $10^{39}$ as in \cite{Miller_2023}, the moment of inertia was sampled from a Gaussian distribution peaked at 1.425 as shown in Fig. \ref{fig:MSP_I_zz_dist} in analogy with the distribution found in Lim et al. \cite{lim2022neutronstarradiideformabilities}. 

\section{Generating the Gravitational Wave Signal} \label{Generating the Gravitational Wave Signal}

\begin{figure}[!t]
    \centering
    \mbox{\includegraphics[width=0.5\textwidth] {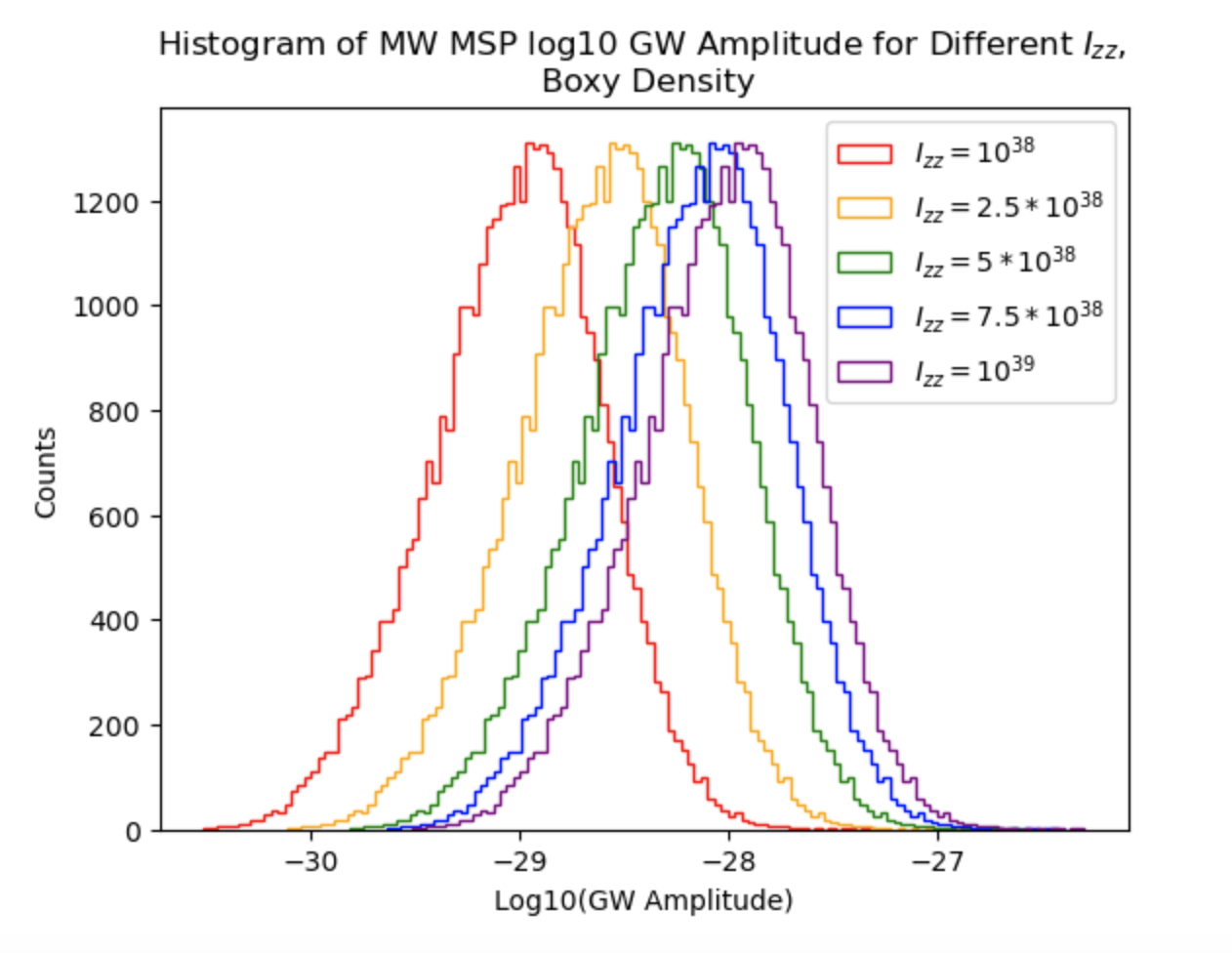}\quad\includegraphics[width=0.52\textwidth]{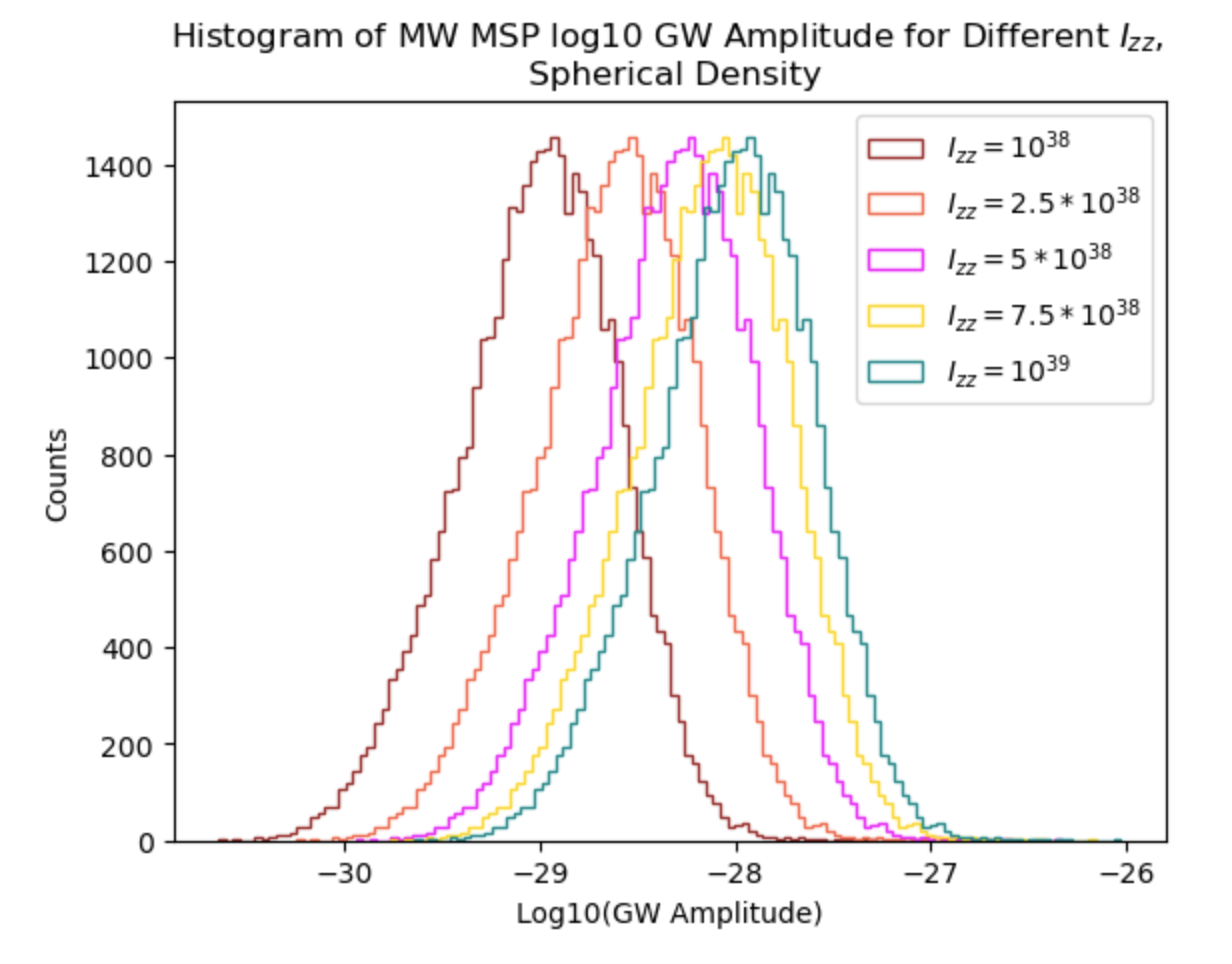}}
    \caption{A histogram of the log10 GW amplitude of boxy (left) and spherical (right) MW sampled MSPs.}
    \label{fig:MSP_GW_amp_hist_log10}
\end{figure}

Calculating the GW amplitude using Eq.~(\ref{GW_amplitude_eqn}) requires a MSP rotational frequency distribution. This was derived from the ATNF Pulsar Catalog, comprising around 800 pulsars with frequency observed via radio and x-ray observations by the ATNF pulsar collaboration \cite{Manchester_2005}. For effective sampling, the frequency distribution was modeled using a Gaussian distribution with $\mu = 288.2$ and $\sigma = 121.4$ Hz. MSPs were assumed to have a frequency of over 100 Hz following Ref.~\cite{Miller_2023}. The resultant GW amplitudes for the two distributions are plotted for a range of moments of inertia $I_{zz}$ in Fig.~\ref{fig:MSP_GW_amp_hist_log10} on a log10 scale. 

\begin{figure}[!t]
    \centering
    \includegraphics[width=0.5\textwidth]{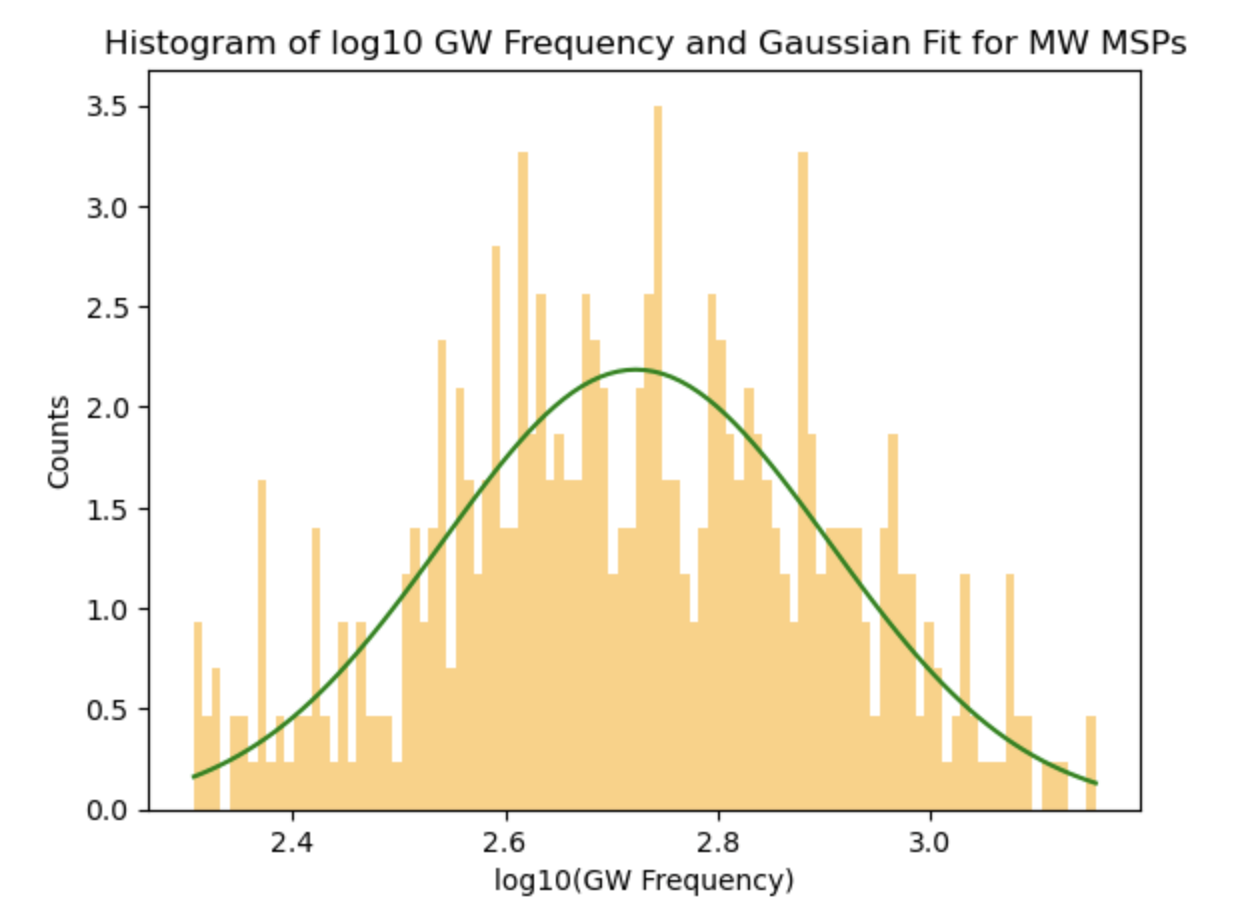}
    \caption{A histogram of the log10 GW frequency of MW sampled MSPs and the Gaussian fit in green. Most MSPs have a GW frequency in the range of $\sim$ 200 to 1400 Hz.}
    \label{fig:MSP_GW_freq_hist}
\end{figure}

The gravitational wave frequency of the MSPs can be calculated using $f_{GW} = 2 f_{rot}$, as in Miller et al. \cite{Miller_2023}, resulting in the distribution shown in Fig.~\ref{fig:MSP_GW_freq_hist}. A normal distribution fit to the data suggested an average log10 frequency of 2.72 Hz, with a width of 0.18 Hz.

\begin{figure}[!t]
    \centering
    \mbox{\includegraphics[width=0.5\textwidth]{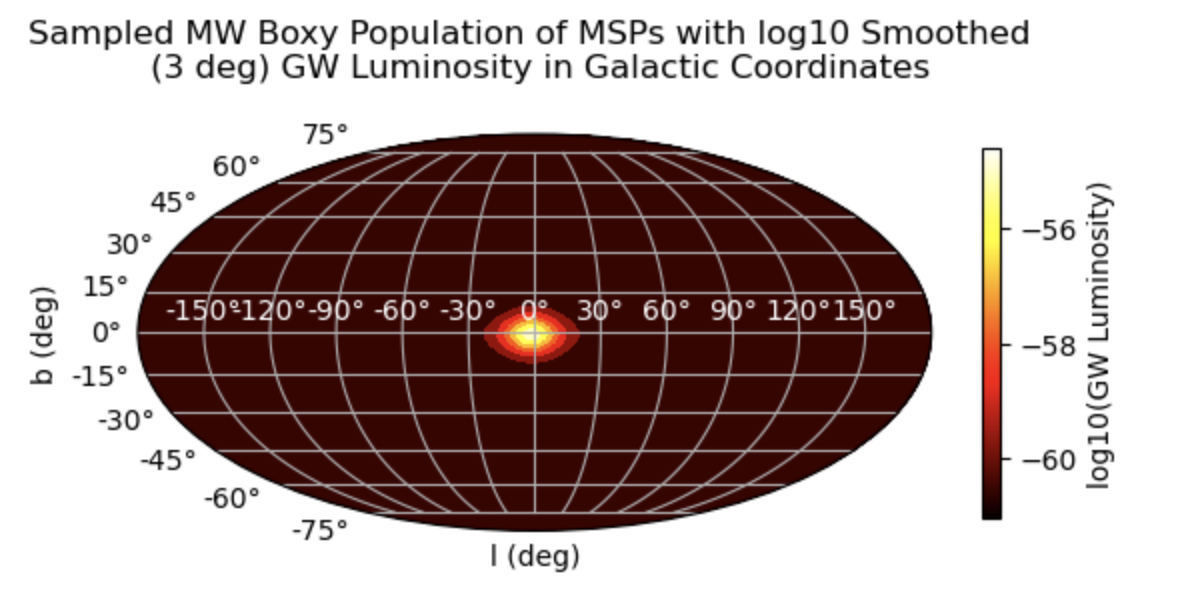}\quad \includegraphics[width=0.52\textwidth]{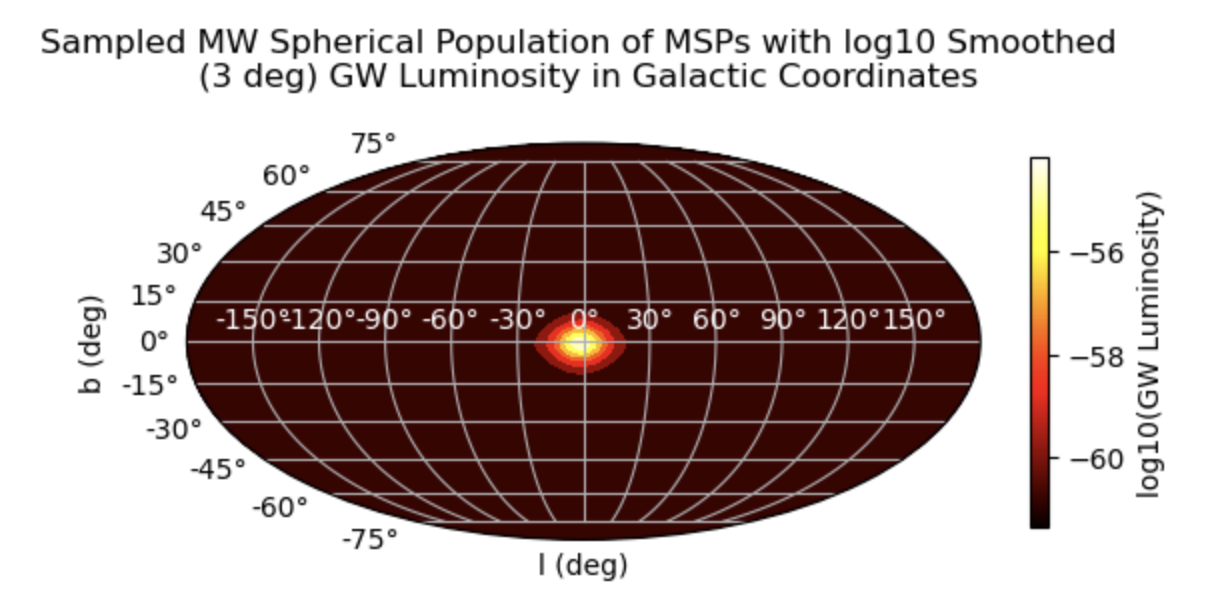}}
    \caption{A Mollweide projection of a sampled of boxy (left) and spherical (right) MW MSP populations with the corresponding log10 GW luminosity, smoothed with an angular resolution of 3 degrees.}
    \label{fig:MSP_GW_smoothed_luminosity_molweide}
\end{figure}

Squaring the GW amplitude results in the GW luminosity; this was then smoothed using an angular resolution of 3 degrees, reflecting the poor angular resolution of most GW detectors \cite{Baker:2019ync}, to obtain the morphology of the signal for both boxy and spherical populations. The corresponding morphology can be seen in Fig.~\ref{fig:MSP_GW_smoothed_luminosity_molweide}. For both populations the morphology of signal parallels that of the MSP distribution, being either boxy or spherical, while decaying significantly due to the high angular precision of 3 degrees.

\section{Analysis and Possibility of Detection} \label{Analysis and Possibility of Detection}

Our results indicate that the gravitational wave frequency of a population of approximately 40,000 MSPs ranges from 2.3 to 3.5 in log space, or approximately 200 to 1400 Hz, owing to the relatively high rotational frequency of most MSPs. This limits the detectability of our signal to GW detectors capable of probing high frequency signals. Comparing our results with the projected sensitivities of most current and planned GW observatories, it is apparent that only detectors such as aLIGO \cite{Buikema_2020}, LIGO A+ \cite{Cahillane_2022}, Advanced Virgo \cite{nguyen2021statusadvancedvirgogravitationalwave}, Kagra \cite{Michimura_2020}, and ET \cite{Hild_2011} are capable of detecting such a signal. The most sensitive, ET, is capable of detecting GW signals above $\sim 10^{-25}$ \cite{Hild_2011} with LIGO A+, aLIGO, Advanced Virgo, and Kagra being capable of detecting GW signals above $\sim 10^{-23}$ \cite{Cahillane_2022, Buikema_2020, nguyen2021statusadvancedvirgogravitationalwave, Michimura_2020}.

Fig.~\ref{fig:MSP_GW_amp_hist_log10} indicates that all GW amplitude values are below $\sim 10^{-26}$, suggesting that both individual MSPs and the collective spectrum, calculated by blurring each individual GW signal with a Gaussian and summing over all ``blurred'' MSPs in each population, would be undetectable. 

\begin{figure}[!t]
    \centering
    \mbox{\includegraphics[width=0.48\textwidth]{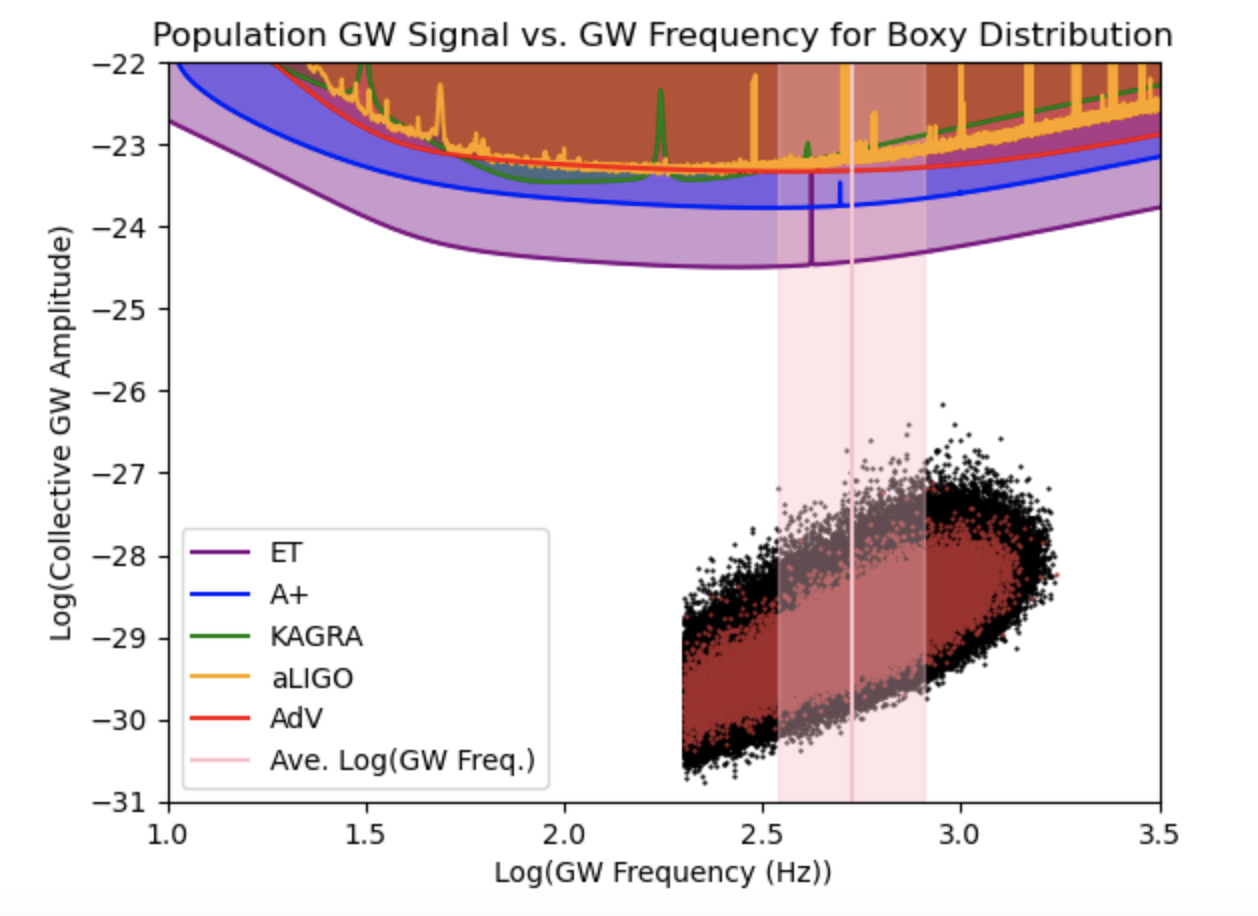}\quad \includegraphics[width=0.5\textwidth]{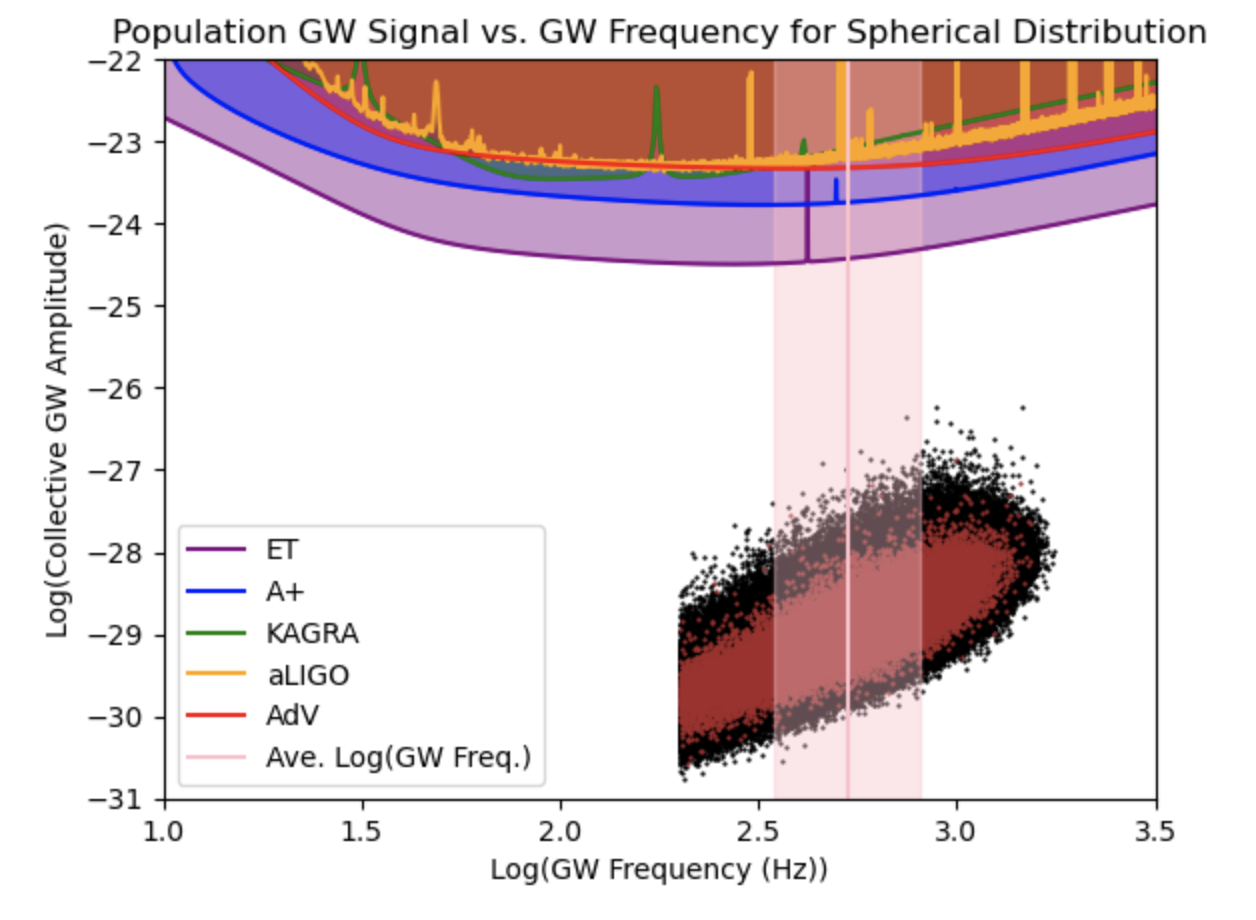}
    }
    \caption{A comparison between the sensitivity curves of current and future GW detectors and the GW signal of individual MSPs of the boxy (left) and spherical (right) population of size 40,000 (brown) and of size 4,000,000 (black) for one generation. The average GW frequency of the MSPs identified in the ATNF pulsar catalog and a one sigma range is plotted in pink for comparison. Note that the truncation in frequency is due to the assumption that MSPs have $f_{rot} > 100$ Hz.}
    \label{fig:MSP_population_GW_vsense}
\end{figure}

In Figure \ref{fig:MSP_population_GW_vsense} we consider both the GW signal for individual MSPs and the collective GW signal versus the corresponding GW frequencies, for both spherical and boxy populations, and compared to the sensitivity curves of aLIGO \cite{Buikema_2020}, LIGO A+ \cite{Cahillane_2022}, Advanced Virgo \cite{nguyen2021statusadvancedvirgogravitationalwave}, Kagra \cite{Michimura_2020}, and ET \cite{Hild_2011}. We show the sample of 40,000 MSPs in brown, and the significantly larger sample of $4\times 10^6$ MSPs in black. The latter is a proxy for a population of extremely {\em gamma-ray dim} MSPs.

\begin{figure}[h!]
    \centering
    \mbox{\includegraphics[width=0.485\textwidth]{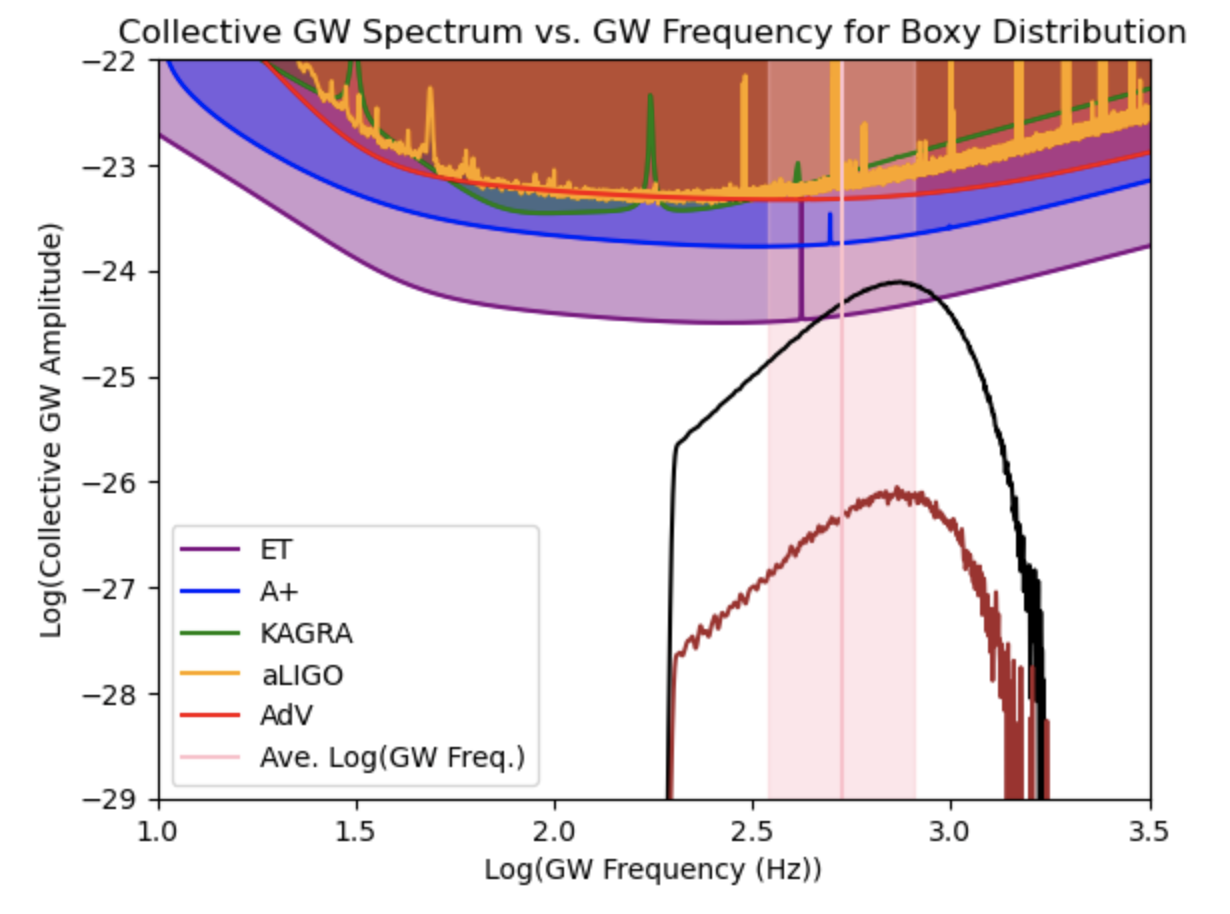}\quad \includegraphics[width=0.5\textwidth]{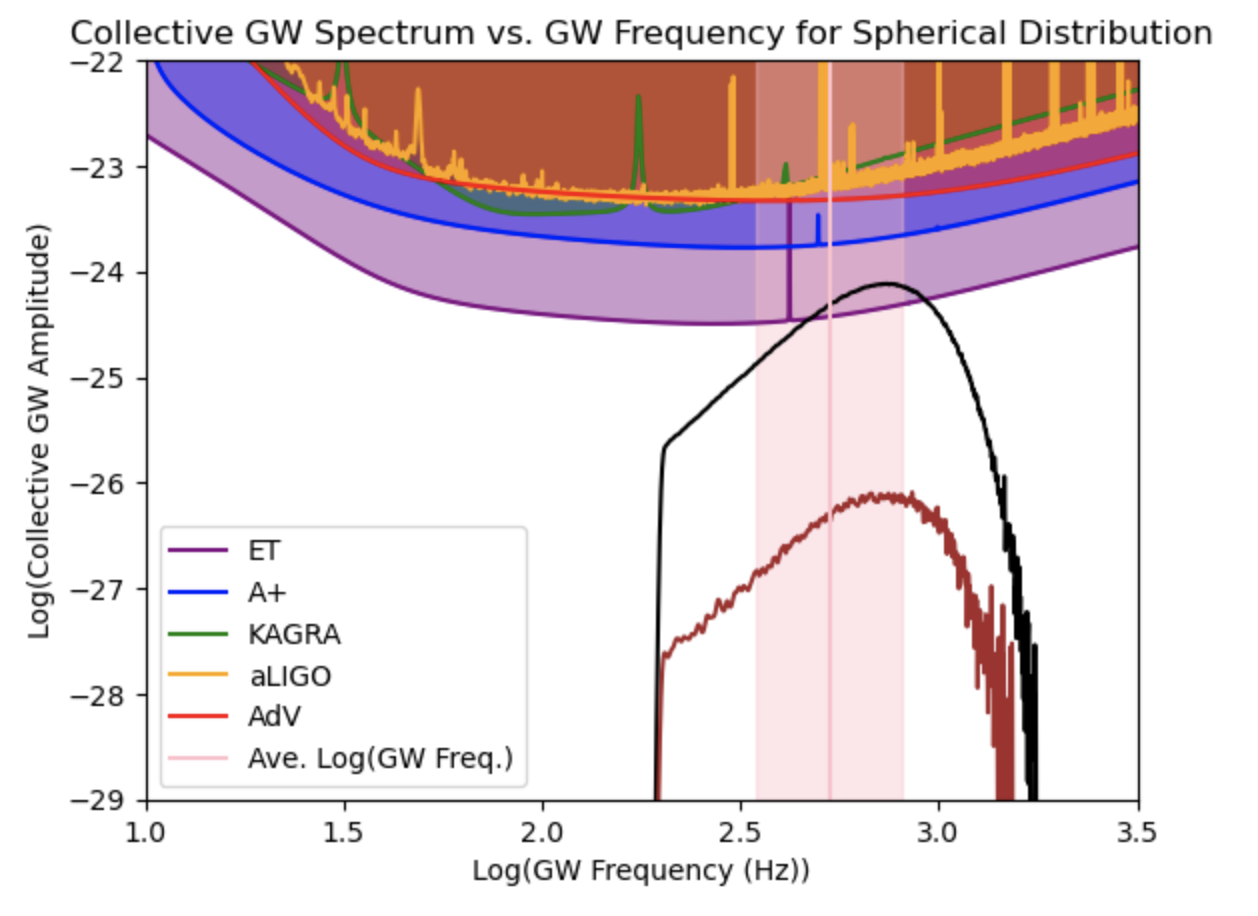}
    }
    \caption{A comparison between the sensitivity curves of current and future GW detectors and the GW spectrum of the boxy (left) and spherical (right) population of size 40,000 (brown) and of size 4,000,000 (black) for one generation. The average GW frequency of the MSPs identified in the ATNF pulsar catalog and a one sigma range is plotted in pink for comparison. Note that the truncation in frequency is due to the assumption that MSPs have $f_{rot} > 100$ Hz.}
    \label{fig:MSP_spectrum_GW_vsense}
\end{figure}

We note that the MSP population in Fig.~\ref{fig:MSP_population_GW_vsense}, right, is significantly more correlated, resulting from the spherical MSP population having less spatial extent than the boxy MSP population. Despite this, both the spherical and boxy populations display GW signals between -30.5 and -27 in log space as required by Fig.~\ref{fig:MSP_GW_amp_hist_log10}. 

This behavior continues in the GW spectra for both populations, with a maximum GW signal of around -26 for both populations. The increase of the maximum GW signal for both populations by one order of magnitude is attributable to the spectrum being representative of the collective signal due to the Gaussian blurring imposed. This renders both the GW signal from individual pulsars and the collective GW spectra undetectable by all current and near-future GW detectors. However, the luminosity function of MSPs is not well-known. If the luminosity function is a factor of 100 smaller than that assumed in \cite{holst2024new} and other studies of the GCE, a much larger population size by a factor of 100 than the assumed 40,000 will be required. Therefore, for a more optimistic outlook, the GW signal for individual MSPs and the GW spectrum for both morphologies were also plotted for a population size of 4,000,000 as seen in black in Fig.~\ref{fig:MSP_population_GW_vsense}. This suggests that the GW signal could be detectable by ET for both morphologies with a larger population size, allowing for constraints to be placed on both the cause of the GCE and the size of the constituent MSP population.

We note that if the GCE is produced by MSPs, a significant fraction thereof would exist within binary systems, as discussed in Sec. \ref{sec:introduction}, and therefore our assumptions that our population of MSPs are isolated or occur with no orbital eccentricity is a significant simplification. Depending on the orbital eccentricity of these binary systems, the MSP's companion star could exert a significant enough tidal force to cause further deformations to the MSP, in addition to the magnetic field deformations already discussed. These could, in turn, significantly modify the accompanying GW amplitudes and frequencies, producing different results than discussed above. This may also necessitate a comparison to different gravitational wave detector sensitivities, as the detectors' sensitivities  in Fig. \ref{fig:MSP_population_GW_vsense} and \ref{fig:MSP_spectrum_GW_vsense} correspond to the case of continuous gravitational wave sources from isolated objects, not objects in binary systems.

\section{Conclusion} \label{Conclusion}

In the age of multi-messenger astrophysics, gravitational wave observatories have the ability to further the understanding of astrophysical phenomenon in new ways. This new avenue of investigation into the universe has to potential to shed light on one of the enduring mysteries of high energy astrophysics: the Galactic center gamma-ray excess. One possible scenario is that this source of gamma rays could be due entirely or in part to a yet unresolved population of MSPs located in the Milky Way disk and the nuclear bulge. Due to their rapid rotation and strong magnetic fields, MSPs are asymmetric about their axis of rotation, causing the emission of monochromatic gravitational waves and allowing the cause of the GCE to potentially be determined. 

Using MCMC and Inverse Transform methods, we generated simulated populations of MSPs and their corresponding ellipticities, moments of inertia, GW amplitude, and GW frequency. Our results indicate that the gravitational wave frequency  ranges from 2.3 to 3.5 in log space, or approximately 200 to 1400 Hz. The GW signal from individual MSPs ranges from $10^{-30.5}$ to $10^{-27}$, and the collective GW spectrum increases to a maximum of $10^{-26}$  for both populations, thus resulting in the signal being undetectable for current and future GW observatories. These conclusions change assuming a dimmer gamma-ray luminosity for individual MSPs and thus a larger population.  

In the future this analysis could be further strengthened by exploring the impact of a time-dependent ellipticity distribution and a frequency-dependent moment of inertia distribution on the strength and detectability of the GW signal. Additionally, it could be that the Galactic center excess is caused by a combination of MSPs and other, more exotic sources such as annihilating dark matter, or other unresolved astrophysical components, so it would be illuminating to incorporate such scenarios into the current model and distribution of MSPs and to evaluate the resulting GW signal. It should also be noted that Holst et al. \cite{holst2024new} suggests a range of possible MSP population sizes from 36,000-47,000 which, along with the unknown MSP luminosity function, suggests that the relation between the GCE and MSP population sizes is relatively poorly constrained. Therefore, further studies into how different population sizes affects detectability would be of great benefit. Finally, an investigation into the effects of binary systems and tidal deformations of MSPs on the resultant gravitational wave signal would give further useful insight into the possibility of using GWs to constrain the cause of the GCE.

\section*{Acknowledgments}

We would like to thank Noe Gonzalez and Emma Strickland for giving their feedback, comments, and advice on this project. This work is partly supported by the U.S.\ Department of Energy grant number de-sc0010107 (SP). 

\bibliography{bib}

\begin{thebibliography}{43}%
\makeatletter
\providecommand \@ifxundefined [1]{%
 \@ifx{#1\undefined}
}%
\providecommand \@ifnum [1]{%
 \ifnum #1\expandafter \@firstoftwo
 \else \expandafter \@secondoftwo
 \fi
}%
\providecommand \@ifx [1]{%
 \ifx #1\expandafter \@firstoftwo
 \else \expandafter \@secondoftwo
 \fi
}%
\providecommand \natexlab [1]{#1}%
\providecommand \enquote  [1]{``#1''}%
\providecommand \bibnamefont  [1]{#1}%
\providecommand \bibfnamefont [1]{#1}%
\providecommand \citenamefont [1]{#1}%
\providecommand \href@noop [0]{\@secondoftwo}%
\providecommand \href [0]{\begingroup \@sanitize@url \@href}%
\providecommand \@href[1]{\@@startlink{#1}\@@href}%
\providecommand \@@href[1]{\endgroup#1\@@endlink}%
\providecommand \@sanitize@url [0]{\catcode `\\12\catcode `\$12\catcode `\&12\catcode `\#12\catcode `\^12\catcode `\_12\catcode `\%12\relax}%
\providecommand \@@startlink[1]{}%
\providecommand \@@endlink[0]{}%
\providecommand \url  [0]{\begingroup\@sanitize@url \@url }%
\providecommand \@url [1]{\endgroup\@href {#1}{\urlprefix }}%
\providecommand \urlprefix  [0]{URL }%
\providecommand \Eprint [0]{\href }%
\providecommand \doibase [0]{http://dx.doi.org/}%
\providecommand \selectlanguage [0]{\@gobble}%
\providecommand \bibinfo  [0]{\@secondoftwo}%
\providecommand \bibfield  [0]{\@secondoftwo}%
\providecommand \translation [1]{[#1]}%
\providecommand \BibitemOpen [0]{}%
\providecommand \bibitemStop [0]{}%
\providecommand \bibitemNoStop [0]{.\EOS\space}%
\providecommand \EOS [0]{\spacefactor3000\relax}%
\providecommand \BibitemShut  [1]{\csname bibitem#1\endcsname}%
\let\auto@bib@innerbib\@empty
\bibitem [{\citenamefont {Goodenough}\ and\ \citenamefont {Hooper}(2009)}]{goodenough2009possibleevidencedarkmatter}%
  \BibitemOpen
  \bibfield  {author} {\bibinfo {author} {\bibfnamefont {L.}~\bibnamefont {Goodenough}}\ and\ \bibinfo {author} {\bibfnamefont {D.}~\bibnamefont {Hooper}},\ }\href {https://arxiv.org/abs/0910.2998} {\  (\bibinfo {year} {2009})},\ \Eprint {http://arxiv.org/abs/0910.2998} {arXiv:0910.2998 [hep-ph]} \BibitemShut {NoStop}%
\bibitem [{\citenamefont {Slatyer}(2022)}]{Slatyer_2022}%
  \BibitemOpen
  \bibfield  {author} {\bibinfo {author} {\bibfnamefont {T.}~\bibnamefont {Slatyer}},\ }\href {\doibase 10.21468/scipostphyslectnotes.53} {\bibfield  {journal} {\bibinfo  {journal} {SciPost Physics Lecture Notes}\ } (\bibinfo {year} {2022}),\ 10.21468/scipostphyslectnotes.53}\BibitemShut {NoStop}%
\bibitem [{\citenamefont {Navarro}\ \emph {et~al.}(1997)\citenamefont {Navarro}, \citenamefont {Frenk},\ and\ \citenamefont {White}}]{Navarro:1996gj}%
  \BibitemOpen
  \bibfield  {author} {\bibinfo {author} {\bibfnamefont {J.~F.}\ \bibnamefont {Navarro}}, \bibinfo {author} {\bibfnamefont {C.~S.}\ \bibnamefont {Frenk}}, \ and\ \bibinfo {author} {\bibfnamefont {S.~D.~M.}\ \bibnamefont {White}},\ }\href {\doibase 10.1086/304888} {\bibfield  {journal} {\bibinfo  {journal} {Astrophys. J.}\ }\textbf {\bibinfo {volume} {490}},\ \bibinfo {pages} {493} (\bibinfo {year} {1997})},\ \Eprint {http://arxiv.org/abs/astro-ph/9611107} {arXiv:astro-ph/9611107} \BibitemShut {NoStop}%
\bibitem [{\citenamefont {Hooper}\ and\ \citenamefont {Linden}(2011)}]{Hooper_2011}%
  \BibitemOpen
  \bibfield  {author} {\bibinfo {author} {\bibfnamefont {D.}~\bibnamefont {Hooper}}\ and\ \bibinfo {author} {\bibfnamefont {T.}~\bibnamefont {Linden}},\ }\href {\doibase 10.1103/physrevd.84.123005} {\bibfield  {journal} {\bibinfo  {journal} {Physical Review D}\ }\textbf {\bibinfo {volume} {84}} (\bibinfo {year} {2011}),\ 10.1103/physrevd.84.123005}\BibitemShut {NoStop}%
\bibitem [{\citenamefont {Abazajian}\ and\ \citenamefont {Kaplinghat}(2012)}]{Abazajian_2012}%
  \BibitemOpen
  \bibfield  {author} {\bibinfo {author} {\bibfnamefont {K.~N.}\ \bibnamefont {Abazajian}}\ and\ \bibinfo {author} {\bibfnamefont {M.}~\bibnamefont {Kaplinghat}},\ }\href {\doibase 10.1103/physrevd.86.083511} {\bibfield  {journal} {\bibinfo  {journal} {Physical Review D}\ }\textbf {\bibinfo {volume} {86}} (\bibinfo {year} {2012}),\ 10.1103/physrevd.86.083511}\BibitemShut {NoStop}%
\bibitem [{\citenamefont {Gordon}\ and\ \citenamefont {Macias}(2013)}]{Gordon_2013}%
  \BibitemOpen
  \bibfield  {author} {\bibinfo {author} {\bibfnamefont {C.}~\bibnamefont {Gordon}}\ and\ \bibinfo {author} {\bibfnamefont {O.}~\bibnamefont {Macias}},\ }\href {\doibase 10.1103/physrevd.88.083521} {\bibfield  {journal} {\bibinfo  {journal} {Physical Review D}\ }\textbf {\bibinfo {volume} {88}} (\bibinfo {year} {2013}),\ 10.1103/physrevd.88.083521}\BibitemShut {NoStop}%
\bibitem [{\citenamefont {Daylan}\ \emph {et~al.}(2016)\citenamefont {Daylan}, \citenamefont {Finkbeiner}, \citenamefont {Hooper}, \citenamefont {Linden}, \citenamefont {Portillo}, \citenamefont {Rodd},\ and\ \citenamefont {Slatyer}}]{Daylan_2016}%
  \BibitemOpen
  \bibfield  {author} {\bibinfo {author} {\bibfnamefont {T.}~\bibnamefont {Daylan}}, \bibinfo {author} {\bibfnamefont {D.~P.}\ \bibnamefont {Finkbeiner}}, \bibinfo {author} {\bibfnamefont {D.}~\bibnamefont {Hooper}}, \bibinfo {author} {\bibfnamefont {T.}~\bibnamefont {Linden}}, \bibinfo {author} {\bibfnamefont {S.~K.}\ \bibnamefont {Portillo}}, \bibinfo {author} {\bibfnamefont {N.~L.}\ \bibnamefont {Rodd}}, \ and\ \bibinfo {author} {\bibfnamefont {T.~R.}\ \bibnamefont {Slatyer}},\ }\href {\doibase 10.1016/j.dark.2015.12.005} {\bibfield  {journal} {\bibinfo  {journal} {Physics of the Dark Universe}\ }\textbf {\bibinfo {volume} {12}},\ \bibinfo {pages} {1‚Äì23} (\bibinfo {year} {2016})}\BibitemShut {NoStop}%
\bibitem [{\citenamefont {Eckner}\ \emph {et~al.}(2018)\citenamefont {Eckner}, \citenamefont {Hou}, \citenamefont {Serpico}, \citenamefont {Winter}, \citenamefont {Zaharijas}, \citenamefont {Martin}, \citenamefont {Mauro}, \citenamefont {Mirabal}, \citenamefont {Petrovic}, \citenamefont {Prodanovic},\ and\ \citenamefont {Vandenbroucke}}]{Eckner_2018}%
  \BibitemOpen
  \bibfield  {author} {\bibinfo {author} {\bibfnamefont {C.}~\bibnamefont {Eckner}}, \bibinfo {author} {\bibfnamefont {X.}~\bibnamefont {Hou}}, \bibinfo {author} {\bibfnamefont {P.~D.}\ \bibnamefont {Serpico}}, \bibinfo {author} {\bibfnamefont {M.}~\bibnamefont {Winter}}, \bibinfo {author} {\bibfnamefont {G.}~\bibnamefont {Zaharijas}}, \bibinfo {author} {\bibfnamefont {P.}~\bibnamefont {Martin}}, \bibinfo {author} {\bibfnamefont {M.~d.}\ \bibnamefont {Mauro}}, \bibinfo {author} {\bibfnamefont {N.}~\bibnamefont {Mirabal}}, \bibinfo {author} {\bibfnamefont {J.}~\bibnamefont {Petrovic}}, \bibinfo {author} {\bibfnamefont {T.}~\bibnamefont {Prodanovic}}, \ and\ \bibinfo {author} {\bibfnamefont {J.}~\bibnamefont {Vandenbroucke}},\ }\href {\doibase 10.3847/1538-4357/aac029} {\bibfield  {journal} {\bibinfo  {journal} {The Astrophysical Journal}\ }\textbf {\bibinfo {volume} {862}},\ \bibinfo {pages} {79} (\bibinfo {year} {2018})}\BibitemShut {NoStop}%
\bibitem [{\citenamefont {Ploeg}(2021)}]{ploeg2021galactic}%
  \BibitemOpen
  \bibfield  {author} {\bibinfo {author} {\bibfnamefont {H.}~\bibnamefont {Ploeg}},\ }\href@noop {} {\enquote {\bibinfo {title} {The galactic millisecond pulsar population: Implications for the galactic center excess},}\ } (\bibinfo {year} {2021}),\ \Eprint {http://arxiv.org/abs/2109.08439} {arXiv:2109.08439 [astro-ph.HE]} \BibitemShut {NoStop}%
\bibitem [{\citenamefont {Lorimer}(2008)}]{MSPreview}%
  \BibitemOpen
  \bibfield  {author} {\bibinfo {author} {\bibfnamefont {D.~R.}\ \bibnamefont {Lorimer}},\ }\href {\doibase 10.12942/lrr-2008-8} {\bibfield  {journal} {\bibinfo  {journal} {Living Rev. Rel.}\ }\textbf {\bibinfo {volume} {11}},\ \bibinfo {pages} {8} (\bibinfo {year} {2008})},\ \Eprint {http://arxiv.org/abs/0811.0762} {arXiv:0811.0762 [astro-ph]} \BibitemShut {NoStop}%
\bibitem [{\citenamefont {Gautam}\ \emph {et~al.}(2022)\citenamefont {Gautam}, \citenamefont {Crocker}, \citenamefont {Ferrario}, \citenamefont {Ruiter}, \citenamefont {Ploeg}, \citenamefont {Gordon},\ and\ \citenamefont {Macias}}]{Gautam:2021wqn}%
  \BibitemOpen
  \bibfield  {author} {\bibinfo {author} {\bibfnamefont {A.}~\bibnamefont {Gautam}}, \bibinfo {author} {\bibfnamefont {R.~M.}\ \bibnamefont {Crocker}}, \bibinfo {author} {\bibfnamefont {L.}~\bibnamefont {Ferrario}}, \bibinfo {author} {\bibfnamefont {A.~J.}\ \bibnamefont {Ruiter}}, \bibinfo {author} {\bibfnamefont {H.}~\bibnamefont {Ploeg}}, \bibinfo {author} {\bibfnamefont {C.}~\bibnamefont {Gordon}}, \ and\ \bibinfo {author} {\bibfnamefont {O.}~\bibnamefont {Macias}},\ }\href {\doibase 10.1038/s41550-022-01658-3} {\bibfield  {journal} {\bibinfo  {journal} {Nature Astron.}\ }\textbf {\bibinfo {volume} {6}},\ \bibinfo {pages} {703} (\bibinfo {year} {2022})},\ \Eprint {http://arxiv.org/abs/2106.00222} {arXiv:2106.00222 [astro-ph.HE]} \BibitemShut {NoStop}%
\bibitem [{\citenamefont {Macias}\ \emph {et~al.}(2019)\citenamefont {Macias}, \citenamefont {Horiuchi}, \citenamefont {Kaplinghat}, \citenamefont {Gordon}, \citenamefont {Crocker},\ and\ \citenamefont {Nataf}}]{Macias_2019}%
  \BibitemOpen
  \bibfield  {author} {\bibinfo {author} {\bibfnamefont {O.}~\bibnamefont {Macias}}, \bibinfo {author} {\bibfnamefont {S.}~\bibnamefont {Horiuchi}}, \bibinfo {author} {\bibfnamefont {M.}~\bibnamefont {Kaplinghat}}, \bibinfo {author} {\bibfnamefont {C.}~\bibnamefont {Gordon}}, \bibinfo {author} {\bibfnamefont {R.~M.}\ \bibnamefont {Crocker}}, \ and\ \bibinfo {author} {\bibfnamefont {D.~M.}\ \bibnamefont {Nataf}},\ }\href {\doibase 10.1088/1475-7516/2019/09/042} {\bibfield  {journal} {\bibinfo  {journal} {Journal of Cosmology and Astroparticle Physics}\ }\textbf {\bibinfo {volume} {2019}},\ \bibinfo {pages} {042‚Äì042} (\bibinfo {year} {2019})}\BibitemShut {NoStop}%
\bibitem [{\citenamefont {Jiang}\ \emph {et~al.}(2020)\citenamefont {Jiang}, \citenamefont {Wang}, \citenamefont {Chen}, \citenamefont {Li}, \citenamefont {Liu},\ and\ \citenamefont {Gao}}]{Jiang:2019hal}%
  \BibitemOpen
  \bibfield  {author} {\bibinfo {author} {\bibfnamefont {L.}~\bibnamefont {Jiang}}, \bibinfo {author} {\bibfnamefont {N.}~\bibnamefont {Wang}}, \bibinfo {author} {\bibfnamefont {W.-C.}\ \bibnamefont {Chen}}, \bibinfo {author} {\bibfnamefont {X.-D.}\ \bibnamefont {Li}}, \bibinfo {author} {\bibfnamefont {W.-M.}\ \bibnamefont {Liu}}, \ and\ \bibinfo {author} {\bibfnamefont {Z.-F.}\ \bibnamefont {Gao}},\ }\href {\doibase 10.1051/0004-6361/201935132} {\bibfield  {journal} {\bibinfo  {journal} {Astron. Astrophys.}\ }\textbf {\bibinfo {volume} {633}},\ \bibinfo {pages} {A45} (\bibinfo {year} {2020})},\ \Eprint {http://arxiv.org/abs/1911.11275} {arXiv:1911.11275 [astro-ph.HE]} \BibitemShut {NoStop}%
\bibitem [{\citenamefont {Chanlaridis}\ \emph {et~al.}(2024)\citenamefont {Chanlaridis}, \citenamefont {Ohse}, \citenamefont {Antoniadis}, \citenamefont {Blaschke}, \citenamefont {Alvarez-Castillo}, \citenamefont {Danchev}, \citenamefont {Misra},\ and\ \citenamefont {Langer}}]{Chanlaridis:2024rov}%
  \BibitemOpen
  \bibfield  {author} {\bibinfo {author} {\bibfnamefont {S.}~\bibnamefont {Chanlaridis}}, \bibinfo {author} {\bibfnamefont {D.}~\bibnamefont {Ohse}}, \bibinfo {author} {\bibfnamefont {J.}~\bibnamefont {Antoniadis}}, \bibinfo {author} {\bibfnamefont {D.}~\bibnamefont {Blaschke}}, \bibinfo {author} {\bibfnamefont {D.~E.}\ \bibnamefont {Alvarez-Castillo}}, \bibinfo {author} {\bibfnamefont {V.}~\bibnamefont {Danchev}}, \bibinfo {author} {\bibfnamefont {D.}~\bibnamefont {Misra}}, \ and\ \bibinfo {author} {\bibfnamefont {N.}~\bibnamefont {Langer}},\ }\href@noop {} {\  (\bibinfo {year} {2024})},\ \Eprint {http://arxiv.org/abs/2409.04755} {arXiv:2409.04755 [astro-ph.HE]} \BibitemShut {NoStop}%
\bibitem [{\citenamefont {McTier}\ \emph {et~al.}(2020)\citenamefont {McTier}, \citenamefont {Kipping},\ and\ \citenamefont {Johnston}}]{McTier2020}%
  \BibitemOpen
  \bibfield  {author} {\bibinfo {author} {\bibfnamefont {M.}~\bibnamefont {McTier}}, \bibinfo {author} {\bibfnamefont {D.}~\bibnamefont {Kipping}}, \ and\ \bibinfo {author} {\bibfnamefont {K.}~\bibnamefont {Johnston}},\ }\href {\doibase 10.1093/mnras/staa1232} {\bibfield  {journal} {\bibinfo  {journal} {Monthly Notices of the Royal Astronomical Society}\ }\textbf {\bibinfo {volume} {495}},\ \bibinfo {pages} {2105} (\bibinfo {year} {2020})}\BibitemShut {NoStop}%
\bibitem [{\citenamefont {Macias}\ \emph {et~al.}(2018)\citenamefont {Macias}, \citenamefont {Gordon}, \citenamefont {Crocker}, \citenamefont {Coleman}, \citenamefont {Paterson}, \citenamefont {Horiuchi},\ and\ \citenamefont {Pohl}}]{Macias_2018}%
  \BibitemOpen
  \bibfield  {author} {\bibinfo {author} {\bibfnamefont {O.}~\bibnamefont {Macias}}, \bibinfo {author} {\bibfnamefont {C.}~\bibnamefont {Gordon}}, \bibinfo {author} {\bibfnamefont {R.~M.}\ \bibnamefont {Crocker}}, \bibinfo {author} {\bibfnamefont {B.}~\bibnamefont {Coleman}}, \bibinfo {author} {\bibfnamefont {D.}~\bibnamefont {Paterson}}, \bibinfo {author} {\bibfnamefont {S.}~\bibnamefont {Horiuchi}}, \ and\ \bibinfo {author} {\bibfnamefont {M.}~\bibnamefont {Pohl}},\ }\href {\doibase 10.1038/s41550-018-0414-3} {\bibfield  {journal} {\bibinfo  {journal} {Nature Astronomy}\ }\textbf {\bibinfo {volume} {2}},\ \bibinfo {pages} {387‚Äì392} (\bibinfo {year} {2018})}\BibitemShut {NoStop}%
\bibitem [{\citenamefont {Storm}\ \emph {et~al.}(2017)\citenamefont {Storm}, \citenamefont {Weniger},\ and\ \citenamefont {Calore}}]{Storm_2017}%
  \BibitemOpen
  \bibfield  {author} {\bibinfo {author} {\bibfnamefont {E.}~\bibnamefont {Storm}}, \bibinfo {author} {\bibfnamefont {C.}~\bibnamefont {Weniger}}, \ and\ \bibinfo {author} {\bibfnamefont {F.}~\bibnamefont {Calore}},\ }\href {\doibase 10.1088/1475-7516/2017/08/022} {\bibfield  {journal} {\bibinfo  {journal} {Journal of Cosmology and Astroparticle Physics}\ }\textbf {\bibinfo {volume} {2017}},\ \bibinfo {pages} {022‚Äì022} (\bibinfo {year} {2017})}\BibitemShut {NoStop}%
\bibitem [{\citenamefont {Buschmann}\ \emph {et~al.}(2020)\citenamefont {Buschmann}, \citenamefont {Rodd}, \citenamefont {Safdi}, \citenamefont {Chang}, \citenamefont {Mishra-Sharma}, \citenamefont {Lisanti},\ and\ \citenamefont {Macias}}]{Buschmann_2020}%
  \BibitemOpen
  \bibfield  {author} {\bibinfo {author} {\bibfnamefont {M.}~\bibnamefont {Buschmann}}, \bibinfo {author} {\bibfnamefont {N.~L.}\ \bibnamefont {Rodd}}, \bibinfo {author} {\bibfnamefont {B.~R.}\ \bibnamefont {Safdi}}, \bibinfo {author} {\bibfnamefont {L.~J.}\ \bibnamefont {Chang}}, \bibinfo {author} {\bibfnamefont {S.}~\bibnamefont {Mishra-Sharma}}, \bibinfo {author} {\bibfnamefont {M.}~\bibnamefont {Lisanti}}, \ and\ \bibinfo {author} {\bibfnamefont {O.}~\bibnamefont {Macias}},\ }\href {\doibase 10.1103/physrevd.102.023023} {\bibfield  {journal} {\bibinfo  {journal} {Physical Review D}\ }\textbf {\bibinfo {volume} {102}} (\bibinfo {year} {2020}),\ 10.1103/physrevd.102.023023}\BibitemShut {NoStop}%
\bibitem [{\citenamefont {Leane}\ and\ \citenamefont {Slatyer}(2020{\natexlab{a}})}]{Leane_2020}%
  \BibitemOpen
  \bibfield  {author} {\bibinfo {author} {\bibfnamefont {R.~K.}\ \bibnamefont {Leane}}\ and\ \bibinfo {author} {\bibfnamefont {T.~R.}\ \bibnamefont {Slatyer}},\ }\href {\doibase 10.1103/physrevlett.125.121105} {\bibfield  {journal} {\bibinfo  {journal} {Physical Review Letters}\ }\textbf {\bibinfo {volume} {125}} (\bibinfo {year} {2020}{\natexlab{a}}),\ 10.1103/physrevlett.125.121105}\BibitemShut {NoStop}%
\bibitem [{\citenamefont {Leane}\ and\ \citenamefont {Slatyer}(2020{\natexlab{b}})}]{Leane_2020_2}%
  \BibitemOpen
  \bibfield  {author} {\bibinfo {author} {\bibfnamefont {R.~K.}\ \bibnamefont {Leane}}\ and\ \bibinfo {author} {\bibfnamefont {T.~R.}\ \bibnamefont {Slatyer}},\ }\href {\doibase 10.1103/physrevd.102.063019} {\bibfield  {journal} {\bibinfo  {journal} {Physical Review D}\ }\textbf {\bibinfo {volume} {102}} (\bibinfo {year} {2020}{\natexlab{b}}),\ 10.1103/physrevd.102.063019}\BibitemShut {NoStop}%
\bibitem [{\citenamefont {Agarwal}\ \emph {et~al.}(2022)\citenamefont {Agarwal}, \citenamefont {Suresh}, \citenamefont {Mandic}, \citenamefont {Matas},\ and\ \citenamefont {Regimbau}}]{Agarwal_2022}%
  \BibitemOpen
  \bibfield  {author} {\bibinfo {author} {\bibfnamefont {D.}~\bibnamefont {Agarwal}}, \bibinfo {author} {\bibfnamefont {J.}~\bibnamefont {Suresh}}, \bibinfo {author} {\bibfnamefont {V.}~\bibnamefont {Mandic}}, \bibinfo {author} {\bibfnamefont {A.}~\bibnamefont {Matas}}, \ and\ \bibinfo {author} {\bibfnamefont {T.}~\bibnamefont {Regimbau}},\ }\href {\doibase 10.1103/PhysRevD.106.043019} {\bibfield  {journal} {\bibinfo  {journal} {Phys. Rev. D}\ }\textbf {\bibinfo {volume} {106}},\ \bibinfo {pages} {043019} (\bibinfo {year} {2022})},\ \Eprint {http://arxiv.org/abs/2204.08378} {arXiv:2204.08378 [gr-qc]} \BibitemShut {NoStop}%
\bibitem [{\citenamefont {Calore}\ \emph {et~al.}(2019)\citenamefont {Calore}, \citenamefont {Regimbau},\ and\ \citenamefont {Serpico}}]{Calore_2019}%
  \BibitemOpen
  \bibfield  {author} {\bibinfo {author} {\bibfnamefont {F.}~\bibnamefont {Calore}}, \bibinfo {author} {\bibfnamefont {T.}~\bibnamefont {Regimbau}}, \ and\ \bibinfo {author} {\bibfnamefont {P.~D.}\ \bibnamefont {Serpico}},\ }\href {\doibase 10.1103/physrevlett.122.081103} {\bibfield  {journal} {\bibinfo  {journal} {Physical Review Letters}\ }\textbf {\bibinfo {volume} {122}} (\bibinfo {year} {2019}),\ 10.1103/physrevlett.122.081103}\BibitemShut {NoStop}%
\bibitem [{\citenamefont {Miller}\ and\ \citenamefont {Zhao}(2023)}]{Miller_2023}%
  \BibitemOpen
  \bibfield  {author} {\bibinfo {author} {\bibfnamefont {A.~L.}\ \bibnamefont {Miller}}\ and\ \bibinfo {author} {\bibfnamefont {Y.}~\bibnamefont {Zhao}},\ }\href {\doibase 10.1103/PhysRevLett.131.081401} {\bibfield  {journal} {\bibinfo  {journal} {Phys. Rev. Lett.}\ }\textbf {\bibinfo {volume} {131}},\ \bibinfo {pages} {081401} (\bibinfo {year} {2023})},\ \Eprint {http://arxiv.org/abs/2301.10239} {arXiv:2301.10239 [astro-ph.HE]} \BibitemShut {NoStop}%
\bibitem [{\citenamefont {Carleo}\ and\ \citenamefont {Ben-Salem}(2023)}]{Carleo:2023qxu}%
  \BibitemOpen
  \bibfield  {author} {\bibinfo {author} {\bibfnamefont {A.}~\bibnamefont {Carleo}}\ and\ \bibinfo {author} {\bibfnamefont {B.}~\bibnamefont {Ben-Salem}},\ }\href {\doibase 10.1103/PhysRevD.108.124027} {\bibfield  {journal} {\bibinfo  {journal} {Phys. Rev. D}\ }\textbf {\bibinfo {volume} {108}},\ \bibinfo {pages} {124027} (\bibinfo {year} {2023})},\ \Eprint {http://arxiv.org/abs/2305.08274} {arXiv:2305.08274 [gr-qc]} \BibitemShut {NoStop}%
\bibitem [{\citenamefont {Abbott}\ \emph {et~al.}(2017)\citenamefont {Abbott} \emph {et~al.}}]{Abbott_2017}%
  \BibitemOpen
  \bibfield  {author} {\bibinfo {author} {\bibfnamefont {B.~P.}\ \bibnamefont {Abbott}} \emph {et~al.} (\bibinfo {collaboration} {LIGO Scientific, Virgo}),\ }\href {\doibase 10.1103/PhysRevLett.119.161101} {\bibfield  {journal} {\bibinfo  {journal} {Phys. Rev. Lett.}\ }\textbf {\bibinfo {volume} {119}},\ \bibinfo {pages} {161101} (\bibinfo {year} {2017})},\ \Eprint {http://arxiv.org/abs/1710.05832} {arXiv:1710.05832 [gr-qc]} \BibitemShut {NoStop}%
\bibitem [{\citenamefont {Carroll}(2019)}]{Carroll:2004st}%
  \BibitemOpen
  \bibfield  {author} {\bibinfo {author} {\bibfnamefont {S.~M.}\ \bibnamefont {Carroll}},\ }\href {\doibase 10.1017/9781108770385} {\emph {\bibinfo {title} {{Spacetime and Geometry}: {An Introduction to General Relativity}}}}\ (\bibinfo  {publisher} {Cambridge University Press},\ \bibinfo {year} {2019})\BibitemShut {NoStop}%
\bibitem [{\citenamefont {Sieniawska}\ and\ \citenamefont {Jones}(2021)}]{Sieniawska_2021}%
  \BibitemOpen
  \bibfield  {author} {\bibinfo {author} {\bibfnamefont {M.}~\bibnamefont {Sieniawska}}\ and\ \bibinfo {author} {\bibfnamefont {D.~I.}\ \bibnamefont {Jones}},\ }\href {\doibase 10.1093/mnras/stab3315} {\bibfield  {journal} {\bibinfo  {journal} {Monthly Notices of the Royal Astronomical Society}\ }\textbf {\bibinfo {volume} {509}},\ \bibinfo {pages} {5179‚Äì5187} (\bibinfo {year} {2021})}\BibitemShut {NoStop}%
\bibitem [{\citenamefont {Holst}\ and\ \citenamefont {Hooper}(2024)}]{holst2024new}%
  \BibitemOpen
  \bibfield  {author} {\bibinfo {author} {\bibfnamefont {I.}~\bibnamefont {Holst}}\ and\ \bibinfo {author} {\bibfnamefont {D.}~\bibnamefont {Hooper}},\ }\href@noop {} {\enquote {\bibinfo {title} {A new determination of the millisecond pulsar gamma-ray luminosity function and implications for the galactic center gamma-ray excess},}\ } (\bibinfo {year} {2024}),\ \Eprint {http://arxiv.org/abs/2403.00978} {arXiv:2403.00978 [astro-ph.HE]} \BibitemShut {NoStop}%
\bibitem [{\citenamefont {Gonthier}\ \emph {et~al.}(2018)\citenamefont {Gonthier}, \citenamefont {Harding}, \citenamefont {Ferrara}, \citenamefont {Frederick}, \citenamefont {Mohr},\ and\ \citenamefont {Koh}}]{refpopsynth}%
  \BibitemOpen
  \bibfield  {author} {\bibinfo {author} {\bibfnamefont {P.~L.}\ \bibnamefont {Gonthier}}, \bibinfo {author} {\bibfnamefont {A.~K.}\ \bibnamefont {Harding}}, \bibinfo {author} {\bibfnamefont {E.~C.}\ \bibnamefont {Ferrara}}, \bibinfo {author} {\bibfnamefont {S.~E.}\ \bibnamefont {Frederick}}, \bibinfo {author} {\bibfnamefont {V.~E.}\ \bibnamefont {Mohr}}, \ and\ \bibinfo {author} {\bibfnamefont {Y.-M.}\ \bibnamefont {Koh}},\ }\href {\doibase 10.3847/1538-4357/aad08d} {\bibfield  {journal} {\bibinfo  {journal} {Astrophys. J.}\ }\textbf {\bibinfo {volume} {863}},\ \bibinfo {pages} {199} (\bibinfo {year} {2018})},\ \Eprint {http://arxiv.org/abs/1806.11215} {arXiv:1806.11215 [astro-ph.HE]} \BibitemShut {NoStop}%
\bibitem [{\citenamefont {Yuan}\ and\ \citenamefont {Zhang}(2014)}]{Yuan_2014}%
  \BibitemOpen
  \bibfield  {author} {\bibinfo {author} {\bibfnamefont {Q.}~\bibnamefont {Yuan}}\ and\ \bibinfo {author} {\bibfnamefont {B.}~\bibnamefont {Zhang}},\ }\href {\doibase 10.1016/j.jheap.2014.06.001} {\bibfield  {journal} {\bibinfo  {journal} {Journal of High Energy Astrophysics}\ }\textbf {\bibinfo {volume} {3‚Äì4}},\ \bibinfo {pages} {1‚Äì8} (\bibinfo {year} {2014})}\BibitemShut {NoStop}%
\bibitem [{\citenamefont {Ploeg}\ \emph {et~al.}(2020)\citenamefont {Ploeg}, \citenamefont {Gordon}, \citenamefont {Crocker},\ and\ \citenamefont {Macias}}]{Ploeg_2020}%
  \BibitemOpen
  \bibfield  {author} {\bibinfo {author} {\bibfnamefont {H.}~\bibnamefont {Ploeg}}, \bibinfo {author} {\bibfnamefont {C.}~\bibnamefont {Gordon}}, \bibinfo {author} {\bibfnamefont {R.}~\bibnamefont {Crocker}}, \ and\ \bibinfo {author} {\bibfnamefont {O.}~\bibnamefont {Macias}},\ }\href {\doibase 10.1088/1475-7516/2020/12/035} {\bibfield  {journal} {\bibinfo  {journal} {Journal of Cosmology and Astroparticle Physics}\ }\textbf {\bibinfo {volume} {2020}},\ \bibinfo {pages} {035–035} (\bibinfo {year} {2020})}\BibitemShut {NoStop}%
\bibitem [{\citenamefont {Foreman-Mackey}\ \emph {et~al.}(2013)\citenamefont {Foreman-Mackey}, \citenamefont {Hogg}, \citenamefont {Lang},\ and\ \citenamefont {Goodman}}]{Foreman_Mackey_2013}%
  \BibitemOpen
  \bibfield  {author} {\bibinfo {author} {\bibfnamefont {D.}~\bibnamefont {Foreman-Mackey}}, \bibinfo {author} {\bibfnamefont {D.~W.}\ \bibnamefont {Hogg}}, \bibinfo {author} {\bibfnamefont {D.}~\bibnamefont {Lang}}, \ and\ \bibinfo {author} {\bibfnamefont {J.}~\bibnamefont {Goodman}},\ }\href {\doibase 10.1086/670067} {\bibfield  {journal} {\bibinfo  {journal} {Publications of the Astronomical Society of the Pacific}\ }\textbf {\bibinfo {volume} {125}},\ \bibinfo {pages} {306‚Äì312} (\bibinfo {year} {2013})}\BibitemShut {NoStop}%
\bibitem [{\citenamefont {Woan}\ \emph {et~al.}(2018)\citenamefont {Woan}, \citenamefont {Pitkin}, \citenamefont {Haskell}, \citenamefont {Jones},\ and\ \citenamefont {Lasky}}]{Woan_2018}%
  \BibitemOpen
  \bibfield  {author} {\bibinfo {author} {\bibfnamefont {G.}~\bibnamefont {Woan}}, \bibinfo {author} {\bibfnamefont {M.~D.}\ \bibnamefont {Pitkin}}, \bibinfo {author} {\bibfnamefont {B.}~\bibnamefont {Haskell}}, \bibinfo {author} {\bibfnamefont {D.~I.}\ \bibnamefont {Jones}}, \ and\ \bibinfo {author} {\bibfnamefont {P.~D.}\ \bibnamefont {Lasky}},\ }\href {\doibase 10.3847/2041-8213/aad86a} {\bibfield  {journal} {\bibinfo  {journal} {The Astrophysical Journal Letters}\ }\textbf {\bibinfo {volume} {863}},\ \bibinfo {pages} {L40} (\bibinfo {year} {2018})}\BibitemShut {NoStop}%
\bibitem [{\citenamefont {Chen}(2020)}]{Chen_2020}%
  \BibitemOpen
  \bibfield  {author} {\bibinfo {author} {\bibfnamefont {W.-C.}\ \bibnamefont {Chen}},\ }\href {\doibase 10.1103/physrevd.102.043020} {\bibfield  {journal} {\bibinfo  {journal} {Physical Review D}\ }\textbf {\bibinfo {volume} {102}} (\bibinfo {year} {2020}),\ 10.1103/physrevd.102.043020}\BibitemShut {NoStop}%
\bibitem [{\citenamefont {Worley}\ \emph {et~al.}(2008)\citenamefont {Worley}, \citenamefont {Krastev},\ and\ \citenamefont {Li}}]{worley2008nuclearconstraintsmomentainertia}%
  \BibitemOpen
  \bibfield  {author} {\bibinfo {author} {\bibfnamefont {A.}~\bibnamefont {Worley}}, \bibinfo {author} {\bibfnamefont {P.~G.}\ \bibnamefont {Krastev}}, \ and\ \bibinfo {author} {\bibfnamefont {B.-A.}\ \bibnamefont {Li}},\ }\href {https://arxiv.org/abs/0801.1653} {\enquote {\bibinfo {title} {Nuclear constraints on the momenta of inertia of neutron stars},}\ } (\bibinfo {year} {2008}),\ \Eprint {http://arxiv.org/abs/0801.1653} {arXiv:0801.1653 [astro-ph]} \BibitemShut {NoStop}%
\bibitem [{\citenamefont {Lim}\ and\ \citenamefont {Holt}(2022)}]{lim2022neutronstarradiideformabilities}%
  \BibitemOpen
  \bibfield  {author} {\bibinfo {author} {\bibfnamefont {Y.}~\bibnamefont {Lim}}\ and\ \bibinfo {author} {\bibfnamefont {J.~W.}\ \bibnamefont {Holt}},\ }\href {https://arxiv.org/abs/2204.09000} {\enquote {\bibinfo {title} {Neutron star radii, deformabilities, and moments of inertia from experimental and ab initio theory constraints on the 208pb neutron skin thickness},}\ } (\bibinfo {year} {2022}),\ \Eprint {http://arxiv.org/abs/2204.09000} {arXiv:2204.09000 [nucl-th]} \BibitemShut {NoStop}%
\bibitem [{\citenamefont {Manchester}\ \emph {et~al.}(2005)\citenamefont {Manchester}, \citenamefont {Hobbs}, \citenamefont {Teoh},\ and\ \citenamefont {Hobbs}}]{Manchester_2005}%
  \BibitemOpen
  \bibfield  {author} {\bibinfo {author} {\bibfnamefont {R.~N.}\ \bibnamefont {Manchester}}, \bibinfo {author} {\bibfnamefont {G.~B.}\ \bibnamefont {Hobbs}}, \bibinfo {author} {\bibfnamefont {A.}~\bibnamefont {Teoh}}, \ and\ \bibinfo {author} {\bibfnamefont {M.}~\bibnamefont {Hobbs}},\ }\href {\doibase 10.1086/428488} {\bibfield  {journal} {\bibinfo  {journal} {The Astronomical Journal}\ }\textbf {\bibinfo {volume} {129}},\ \bibinfo {pages} {1993‚Äì2006} (\bibinfo {year} {2005})}\BibitemShut {NoStop}%
\bibitem [{\citenamefont {Baker}\ \emph {et~al.}(2021)\citenamefont {Baker} \emph {et~al.}}]{Baker:2019ync}%
  \BibitemOpen
  \bibfield  {author} {\bibinfo {author} {\bibfnamefont {J.}~\bibnamefont {Baker}} \emph {et~al.},\ }\href {\doibase 10.1007/s10686-021-09712-0} {\bibfield  {journal} {\bibinfo  {journal} {Exper. Astron.}\ }\textbf {\bibinfo {volume} {51}},\ \bibinfo {pages} {1441} (\bibinfo {year} {2021})},\ \Eprint {http://arxiv.org/abs/1908.11410} {arXiv:1908.11410 [astro-ph.HE]} \BibitemShut {NoStop}%
\bibitem [{\citenamefont {Buikema}\ \emph {et~al.}(2020)\citenamefont {Buikema} \emph {et~al.}}]{Buikema_2020}%
  \BibitemOpen
  \bibfield  {author} {\bibinfo {author} {\bibfnamefont {A.}~\bibnamefont {Buikema}} \emph {et~al.} (\bibinfo {collaboration} {aLIGO}),\ }\href {\doibase 10.1103/PhysRevD.102.062003} {\bibfield  {journal} {\bibinfo  {journal} {Phys. Rev. D}\ }\textbf {\bibinfo {volume} {102}},\ \bibinfo {pages} {062003} (\bibinfo {year} {2020})},\ \Eprint {http://arxiv.org/abs/2008.01301} {arXiv:2008.01301 [astro-ph.IM]} \BibitemShut {NoStop}%
\bibitem [{\citenamefont {Cahillane}\ and\ \citenamefont {Mansell}(2022)}]{Cahillane_2022}%
  \BibitemOpen
  \bibfield  {author} {\bibinfo {author} {\bibfnamefont {C.}~\bibnamefont {Cahillane}}\ and\ \bibinfo {author} {\bibfnamefont {G.}~\bibnamefont {Mansell}},\ }\href {\doibase 10.3390/galaxies10010036} {\bibfield  {journal} {\bibinfo  {journal} {Galaxies}\ }\textbf {\bibinfo {volume} {10}},\ \bibinfo {pages} {36} (\bibinfo {year} {2022})}\BibitemShut {NoStop}%
\bibitem [{\citenamefont {Nguyen}(2021)}]{nguyen2021statusadvancedvirgogravitationalwave}%
  \BibitemOpen
  \bibfield  {author} {\bibinfo {author} {\bibfnamefont {C.}~\bibnamefont {Nguyen}},\ }\href {https://arxiv.org/abs/2105.09247} {\enquote {\bibinfo {title} {Status of the advanced virgo gravitational-wave detector},}\ } (\bibinfo {year} {2021}),\ \Eprint {http://arxiv.org/abs/2105.09247} {arXiv:2105.09247 [astro-ph.IM]} \BibitemShut {NoStop}%
\bibitem [{\citenamefont {Michimura}\ \emph {et~al.}(2020)\citenamefont {Michimura}, \citenamefont {Komori}, \citenamefont {Enomoto}, \citenamefont {Nagano}, \citenamefont {Nishizawa}, \citenamefont {Hirose}, \citenamefont {Leonardi}, \citenamefont {Capocasa}, \citenamefont {Aritomi}, \citenamefont {Zhao}, \citenamefont {Flaminio}, \citenamefont {Ushiba}, \citenamefont {Yamada}, \citenamefont {Wei}, \citenamefont {Takeda}, \citenamefont {Tanioka}, \citenamefont {Ando}, \citenamefont {Yamamoto}, \citenamefont {Hayama}, \citenamefont {Haino},\ and\ \citenamefont {Somiya}}]{Michimura_2020}%
  \BibitemOpen
  \bibfield  {author} {\bibinfo {author} {\bibfnamefont {Y.}~\bibnamefont {Michimura}}, \bibinfo {author} {\bibfnamefont {K.}~\bibnamefont {Komori}}, \bibinfo {author} {\bibfnamefont {Y.}~\bibnamefont {Enomoto}}, \bibinfo {author} {\bibfnamefont {K.}~\bibnamefont {Nagano}}, \bibinfo {author} {\bibfnamefont {A.}~\bibnamefont {Nishizawa}}, \bibinfo {author} {\bibfnamefont {E.}~\bibnamefont {Hirose}}, \bibinfo {author} {\bibfnamefont {M.}~\bibnamefont {Leonardi}}, \bibinfo {author} {\bibfnamefont {E.}~\bibnamefont {Capocasa}}, \bibinfo {author} {\bibfnamefont {N.}~\bibnamefont {Aritomi}}, \bibinfo {author} {\bibfnamefont {Y.}~\bibnamefont {Zhao}}, \bibinfo {author} {\bibfnamefont {R.}~\bibnamefont {Flaminio}}, \bibinfo {author} {\bibfnamefont {T.}~\bibnamefont {Ushiba}}, \bibinfo {author} {\bibfnamefont {T.}~\bibnamefont {Yamada}}, \bibinfo {author} {\bibfnamefont {L.-W.}\ \bibnamefont {Wei}}, \bibinfo {author} {\bibfnamefont {H.}~\bibnamefont {Takeda}}, \bibinfo {author} {\bibfnamefont {S.}~\bibnamefont
  {Tanioka}}, \bibinfo {author} {\bibfnamefont {M.}~\bibnamefont {Ando}}, \bibinfo {author} {\bibfnamefont {K.}~\bibnamefont {Yamamoto}}, \bibinfo {author} {\bibfnamefont {K.}~\bibnamefont {Hayama}}, \bibinfo {author} {\bibfnamefont {S.}~\bibnamefont {Haino}}, \ and\ \bibinfo {author} {\bibfnamefont {K.}~\bibnamefont {Somiya}},\ }\href {\doibase 10.1103/physrevd.102.022008} {\bibfield  {journal} {\bibinfo  {journal} {Physical Review D}\ }\textbf {\bibinfo {volume} {102}} (\bibinfo {year} {2020}),\ 10.1103/physrevd.102.022008}\BibitemShut {NoStop}%
\bibitem [{\citenamefont {Hild}\ \emph {et~al.}(2011)\citenamefont {Hild} \emph {et~al.}}]{Hild_2011}%
  \BibitemOpen
  \bibfield  {author} {\bibinfo {author} {\bibfnamefont {S.}~\bibnamefont {Hild}} \emph {et~al.},\ }\href {\doibase 10.1088/0264-9381/28/9/094013} {\bibfield  {journal} {\bibinfo  {journal} {Class. Quant. Grav.}\ }\textbf {\bibinfo {volume} {28}},\ \bibinfo {pages} {094013} (\bibinfo {year} {2011})},\ \Eprint {http://arxiv.org/abs/1012.0908} {arXiv:1012.0908 [gr-qc]} \BibitemShut {NoStop}%
\end{thebibliography}%

\end{document}